\documentclass[aps,pra,showpacs,groupedaddress,superscriptaddress,twocolumn,longbibliography]{revtex4-2}

\usepackage{dsfont}
\usepackage{float}
\usepackage{xcolor}
\usepackage{physics}
\usepackage[caption=false]{subfig}

\usepackage{hyperref}
\hypersetup{
	breaklinks=true,
	colorlinks=true,
	allcolors=blue,
}

\usepackage{txfonts}
\usepackage{amsmath}
\usepackage{amssymb}
\usepackage{graphicx}
\usepackage{dcolumn}
\usepackage{bm}
\usepackage{array}
\usepackage{multirow}

\begin{document}

 \title{User-friendly TWA for dissipative spin dynamics}

	\author{Hossein Hosseinabadi\href{https://orcid.org/0009-0002-3427-132X}{\includegraphics[height=1.7ex]{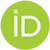}}}
\thanks{These authors contributed equally to this work.}
\affiliation{                    
	Institut f\"{u}r Physik, Johannes Gutenberg Universit\"{a}t Mainz, D-55099 Mainz, Germany
}

	\author{Oksana Chelpanova\href{https://orcid.org/0000-0002-1679-1359}{\includegraphics[height=1.7ex]{orcid-logo.png}}}
\thanks{These authors contributed equally to this work.}
	\affiliation{                    
	Institut f\"{u}r Physik, Johannes Gutenberg Universit\"{a}t Mainz, D-55099 Mainz, Germany
}

	\author{Jamir Marino\href{https://orcid.org/0000-0003-2585-2886}{\includegraphics[height=1.7ex]{orcid-logo.png}}}
    \email{jamirmar@buffalo.edu}
	\affiliation{                    
		Institut f\"{u}r Physik, Johannes Gutenberg Universit\"{a}t Mainz, D-55099 Mainz, Germany
	}

\affiliation{Department of Physics, The State University of New York at Buffalo, Buffalo, NY 14260, USA}

	\begin{abstract}
    We put forward a user-friendly framework of the truncated Wigner approximation (TWA) for dissipative quantum many-body systems. Our approach is computationally affordable and it features a straightforward implementation. The leverage of  the method can be ultimately traced to an intimate connection between the TWA and the semi-classical limit of the quantum Langevin equation, which we unveil by resorting to a path integral representation of the Lindbladian. Our approach allows us to explore dynamics from early to late times in a   variety of models at the core of modern AMO research, including lasing, central spin models, driven arrays of Rydbergs and correlated emission in free space. Notably, our TWA approach outperforms the cumulant expansion method in certain models and performs comparably well in others, all while offering significantly lower computational costs and a much simpler formulation of the dynamical equations. We therefore argue that   TWA could become in the near future a primary   tool  for  a fast and efficient first exploration of  driven-dissipative many-body dynamics on   consumer grade computers.

        
	\end{abstract}
	\maketitle

	\section{Introduction}

	The dynamics of dissipative quantum many-body systems is a central topic of solid-state physics, atomic, molecular, and optical (AMO) physics, as well as quantum information science. Nearly all modern experimental platforms and quantum simulators can be modeled as interacting many-particle systems that exhibit some degree of quantum coherence, are potentially driven by external fields, and are coupled to an environment. 
	Understanding isolated quantum many-body systems is already technically challenging, and becomes further complicated by introducing system-environment coupling. In open systems, dissipation can either suppress quantum phenomena or give rise to novel effects from the interplay of interactions and dissipation.~\cite{fazio2024many,haroche2006exploring,PRXQuantum.3.010201,sieberer2016keldysh,sieberer2023universality,DonnerReview}.

	For a wide range of open quantum systems, it is possible to obtain a time-local description of dynamics that involves only the system's degrees of freedom, without explicitly including those of the environment. In these cases, the system's density matrix evolves according to the Lindblad master equation given by~\cite{Breuer_Petruccione}
	\begin{equation}\label{eq:LindbladMaster}
		\dv{}{t}\hat\rho=-i[\hat H, \hat\rho]+\sum_i \Gamma_{ji}\,\Big(\hat L_i \hat \rho \hat L_j^{\dagger}-\frac{1}{2}\big\{\hat L_j^{\dagger} \hat L_i, \hat\rho \big\}\Big),
	\end{equation}
	where the Hamiltonian $\hat{H}$ governs coherent dynamics, while the second term captures environment-induced dissipation, under the assumption of a completely positive matrix $\Gamma_{ij}$. Essentially, the Lindblad formalism simplifies the study of open-system dynamics by encapsulating dissipation in the jump operators $\hat{L}_i$, reducing the computational resources required to solve for the density matrix, compared to treating the system and environment together.

	However, the numerical costs of solving Eq.~\eqref{eq:LindbladMaster} for generic systems still remain restrictive with current classical computers. Working with the density matrix, rather than pure states as in   unitary dynamics,   leads to a significantly faster growth in computational complexity with system size, which makes exact numerics impractical even for systems of a few atoms and a single photon mode.
    A reduced computational cost can be achieved by unraveling the Lindblad master equation (method of `quantum trajectories'), which bypasses the need to solve for the density matrix~\cite{Qutip,daley2014quantum}. 
Nevertheless, without resorting to approximations,  about a dozen degrees of freedom remains the upper limit to the exact numerical solutions of  Eq.~\eqref{eq:LindbladMaster}.

    Progress in scientific discovery is fueled by  flexible and speedy  approaches to reliably test starting hypotheses, which can then serve as a solid starting point for the development of more sophisticated and resource-intensive methods. In the view of the authors' of this work, such approach  should meet three key criteria. (i) It is computationally affordable, allowing the study of dynamics in the many-body limit and for sufficiently long times using the available classical computers (possibly, even standard consumer laptops). (ii) It captures quantum effects to a sufficient extent. Notably, this requirement is relatively modest for most driven-open many-body AMO (or solid state) systems, as they typically exhibit limited quantum fluctuations due to major role played by dissipation. Many-body phenomena driven by strong quantum effects  are   rare in these platforms and stand out precisely due to their pronounced quantum character (see for instance~\cite{Semeghini_spinliquid2021,chiocchetta2021cavity,JMQSL} for spin liquids in light-matter interfaces). (iii) The method should feature a straightforward implementation, requiring no fine-tuning for each specific problem, and, ideally,  be accessible to first-time users with minimal effort. 

    Our work aims at putting forward a semi-classical method for solving   driven-dissipative quantum many body systems,  satisfying all the three above mentioned criteria. 
    This is a long-standing problem since the inception of the truncated Wigner approximation (TWA) for spin systems~\cite{polkovnikov2010phase,Schachenmayer2015} -- a phase space method suited to capture the semi-classical dynamics of many-body spin systems. Major efforts in extending this approach to dissipative spin systems have culminated in the body of works in Refs. ~\cite{qu2019spin,Liu_PRL2020,Huber_DDTWA,Rabl_phase_space_2021, Fleischhauer_2022,mink2023collective,cascadedSR,Singh_OSDTWA2022}. However, each of them misses some of the aspects identified in criterion (iii), either minimal learning efforts, or broad applicability, holding back the development of a universal semi-classical framework for dissipative dynamics. This remains, however, a high-priority  task, since a method capable to simulate on a personal computer the dynamics of driven-open quantum many body systems in short time, would have a transformative impact for the field, both by offering quick benchmarks to experiments or steering  theory towards the development of more refined techniques.

    To better appreciate the relevance of our framework, we begin by briefly summarizing current computational methods for driven open quantum systems, and subsequently review the state of the art in semi-classical approaches.
    An accessible method, which is widely used as the first attempt to attack new problems, is the mean-field (MF) approximation. By neglecting all quantum corrections, MF describes the system through classical equations of motion for the expectation values of physical observables. This approach is particularly effective for   all-to-all interacting models (such as Dicke, Tavis-Cummings, or Lipkin-Meshkov-Glick), where quantum fluctuations are suppressed as the system size increases~\cite{marino2022dynamical,kirton2017suppressing,Chelpanova_ETH,Chelpanova_Intertwining_2023,kirton2019introduction,defenu}. MF does, therefore, address the criteria (i) and (iii) above, while failing on (ii). Beyond the collective limit, MF is well-known to give inaccurate predictions, and its accuracy  can be improved by accounting for correlations between observables, captured by the connected parts of multi-point correlation functions. This approach leads to an infinite hierarchy of coupled differential equations for correlation functions. Truncating this hierarchy at a finite level, by neglecting higher-order connected correlation functions and obtaining a closed set of equations, results in the method of cumulant expansion (CE)~\cite{kirton2018superradiant,robicheaux2021beyond,Yelin_cumulants,plankensteiner2022quantumcumulants,holzinger2024symmetry}. Usually, CE provides significant improvement over MF results, sometimes even matching the exact solution~\cite{kirton2018superradiant}. However, when truncation at a given order fails to produce satisfactory results, or suffers from numerical instabilities~\cite{Yelin_cumulants}, or when higher-order correlations need to be explored, one must retain additional equations in the hierarchy or selectively exclude certain correlation functions based on heuristic choices. In some cases, the accuracy of CE decreases at higher orders of the expansion~\cite{PhysRevResearch.5.033148,pavskauskas2012equilibration}. These issues introduce an arbitrary element into the approximation, compromising its reliability and, essentially, requiring prior knowledge of the solution to the problem. Therefore, while improving on the criterion (ii) by introducing quantum fluctuations on top of mean-field, it fails criterion (iii), and to some extent  (i), as we will also further discuss in the body of the paper.

    More advanced techniques such as tensor networks ~\cite{ORUS2014,orus_methods,vidal_2003,vidal_2004,Daley_2009MPStrajectories,mc2021stable}, variational principles~\cite{JACKIW1979,Haegeman_TDVP2011,Cui_VPMPO2015} and quantum kinetic equations based on diagrammatics~\cite{Berges_2004,Kamenev_2023,sieberer2023universality,hosseinabadi_YSYK2023,stefanini2024dissipativerealizationkondomodels,qu2024variational,Buchfold_SG2013,hosseinabadi2023dynamics,hosseinabadi2023nonequilibrium,babadi2017theory,lang2024field,chakraborty2022controlling}, require substantial expertise and time investment. While the former is of particular great efficacy in low dimensions, the latter two require strong physical intuition and solid theoretical background on the problem at hand. These methods represent primary examples of those sophisticated and resource-intensive approaches, that should be employed once it becomes clear that solving the physical problem under scrutiny requires the inclusion of strong correlations  on long time scales and large system sizes. As  methods suited for first exploratory studies, they would certainly   fail criteria (i) and (iii).

	For \emph{isolated systems}, TWA is a method that satisfies all the three criteria outlined above. Broadly speaking, it approximates quantum dynamics using classical statistical mechanics, where quantum uncertainty is mapped onto a classical probability distribution through the Wigner transformation of system's density matrix~\cite{polkovnikov2010phase}, and expectation values of observables are approximated by their statistical averages over an ensemble of classical trajectories. In the absence of external dissipation, each trajectory is initialized by sampling from the corresponding probability distribution and evolves according to classical equations of motion. While these equations resemble those of MF, the statistical sampling of initial conditions accounts for leading-order quantum fluctuations (criterion (ii)). TWA has been notably successful in describing the dynamics of isolated bosonic~\cite{Blakie_TWA2008,Polkovnikov_2003} and spin systems~\cite{Schachenmayer2015}, and has been promisingly extended to fermions~\cite{Davidson_FermTWA2017}.  The combination of a MF logic with a straightforward stochastic sampling makes TWA computationally affordable even for large system sizes and long times (i), as well as   easy to implement (iii). 
	
	However, extending TWA to \emph{open quantum systems} has proven more challenging, with success largely limited to some of the simplest dissipation channels.  In general, dissipation arises from the interplay between deterministic and stochastic elements where, analogous to the classical Langevin equation, the deterministic component appears as damping, while the stochastic component manifests as noise~\cite{Kamenev_2023,kardar2007statistical}. The key challenge in dissipative TWA is to properly incorporate both elements into the equations of motion, which has led to various issues in existing implementations. These issues include spin-length shrinkage for individual trajectories, which erases the relevant physics beyond short times~\cite{qu2019spin,Liu_PRL2020,Huber_DDTWA}, and restricted applicability to collective (i.e., very large) spin sizes~\cite{Rabl_phase_space_2021}.
  The mentioned issues can be overcome via a hybrid continuous-discrete Truncated Wigner approximation~\cite{Fleischhauer_2022,mink2023collective,cascadedSR}, or by combining phase space methods with stochastic projections~\cite{Singh_OSDTWA2022}. This variety of approaches calls for a universal framework and naturally raises the question of whether a dissipative TWA can be derived that combines broad applicability with the added advantages of computational efficiency and ease of use. Ideally, such a formulation would be accessible even to users with little to no background in phase-space representations of quantum dynamics, with the potential to accelerate progress in both theoretical and experimental AMO physics.
    
\begin{figure*}
		\centering
		\includegraphics[width=1.0\linewidth]{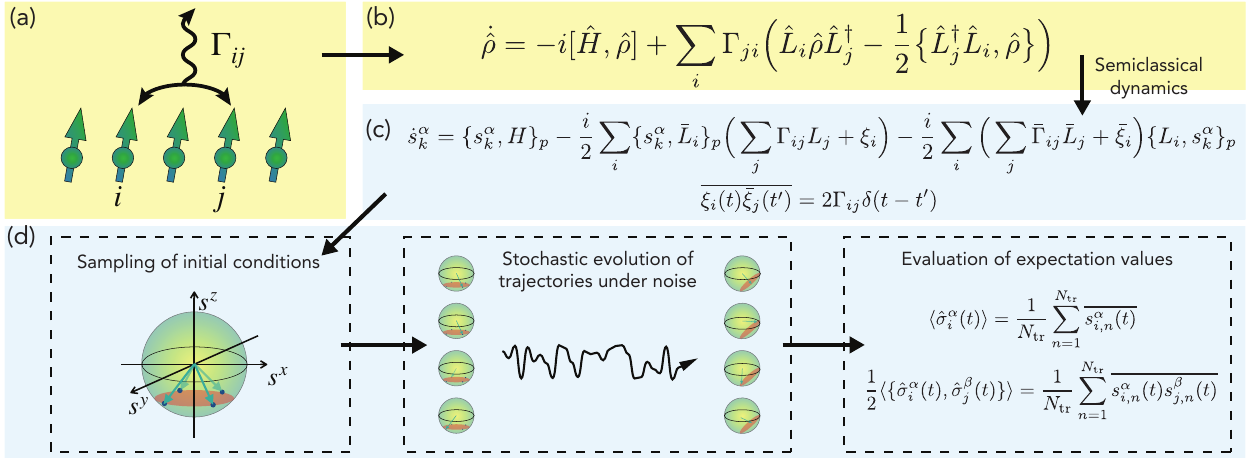}
		\caption{\textbf{Schematic overview of the method for dissipative spins.} (a) A system of spins subjected to dissipation. (b) The exact quantum dynamics are governed by the Lindblad master equation. (c) The semi-classical approximation replaces quantum spins with classical variables, evolving according to stochastic differential equations. (d) Classical spin variables are initialized by sampling from a distribution function  (shown here with discrete sampling) and then evolved under the classical equations with added noise. Quantum expectation values are approximated by averaging over multiple trajectories and noise realizations.}
		\label{fig:schematics}
	\end{figure*}

	In this work, we present a universal and state-of-art formulation of TWA for Lindblad dynamics that addresses all major issues discussed above. Our approach is general and applies to arbitrary  degrees of freedom and dissipation channels. The method has a low entry barrier, as it follows a  recipe that enables to write straightforwardly the equations of motion for a given model. Moreover,  these equations can typically be simulated within minutes on a standard laptop, making the approach both   accessible to a broad community of users and computationally efficient.
    Despite its flexibility, the method remains highly intuitive, requiring only a basic understanding of stochastic differential equations. Due to its systematic derivation, it provides a controlled and conserving approximation, avoiding issues such as artificial spin shrinkage~\cite{Rabl_phase_space_2021}. We develop the formalism using the Keldysh path-integral representation of quantum dynamics, and by expanding the action to second order in quantum fluctuations~\cite{Kamenev_2023}. Our work is inspired by the approach of Refs.~\cite{Polkovnikov_2003,polkovnikov2010phase}, which derived TWA for closed bosonic systems by truncating the Keldysh action at first order in quantum fluctuations. While the derivation relies on field theoretical arguments, the resulting framework can be applied without prior knowledge of field theory, by resorting to a simple  recipe outlined in Sec.~\ref{sec:semiclassics} (see also Fig.~\ref{fig:schematics}). Notably, the equations of motion admit an intuitive interpretation  as a semi-classical limit of the quantum Langevin equation  (QLE)  for dissipative operator dynamics~\cite{gardiner1985handbook}. By applying our method to various examples and comparing the results to exact solutions, we demonstrate its overall advantages over other approaches. In particular, our method surpasses or does comparatively well to CE in all the examples   considered. Thus, we believe that our framework has the potential to become a first-choice tool for studying the dynamics of dissipative quantum many-body systems.

	We begin in Sec.~\ref{sec:method} by introducing a straightforward protocol for applying our dissipative TWA to general problems, while postponing its formal field-theoretic derivation to Appendix~\ref{app:derivation}. In Sec.~\ref{sec:spins}, we specialize our approach to dissipative spin systems before applying it to a series of increasingly complex examples. These include a single driven spin, the Tavis-Cummings model for lasing, the central spin model, a driven Rydberg chain, and finally, the dynamics of a sub-wavelength atomic chain with correlated emission. We conclude in Sec.~\ref{sec:conclusions} by giving an overview of the work, and discussing potential applications of the method and exploring its possible extensions.

	\section{Method\label{sec:method}}

	In Section~\ref{sec:semiclassics} below, we focus on the main result of this work, by expressing the method in terms of an effective classical Hamiltonian that encapsulates all relevant aspects of the approximation. The field-theoretic derivation is presented in detail in Appendix~\ref{app:derivation}, and does not rely on a Fokker-Planck equation for the classical distribution function. Instead, it operates directly with classical trajectories and their averages. In Sec.~\ref{sec:QLE}, we demonstrate that the equations of motion closely resemble QLE for dissipative operator dynamics, with quantum commutators replaced by classical Poisson brackets.

\subsection{General protocol\label{sec:semiclassics}}

	We consider the dynamics of Lindblad systems as given by the general expression in Eq.~\eqref{eq:LindbladMaster}. Taking $\hat{\psi}_\alpha$ as the set of basic operator degrees of freedom in the system, e.g. spins or boson creation and annihilation operators, the semi-classical dynamics can be obtained using the following prescription (see Appendix~\ref{app:derivation} for a derivation):
	\begin{enumerate}
		\item Replace quantum operators $\hat{\psi}_\alpha$ with the classical dynamical variables $\psi_\alpha$ and consequently, obtain the classical Hamiltonian ($\hat{H}\to H$) and jump operators ($\hat{L}_i\to L_i$).
		\item Construct the following effective classical Hamiltonian which captures dissipation by coupling the system's jump variables to a set of self-consistent fields:
		\begin{equation}\label{eq:jm}
			\tilde{H}= H - i \sum_{i} \big(\bar{L}_i \Phi_i - \bar{\Phi}_i L_i \big),
		\end{equation}
        where the bar stands for complex conjugation. This yields   the equations of motion for $\psi_\alpha$ from
		\begin{equation}
			\dv{}{t}\psi_\alpha = \{ \psi_\alpha,\tilde{H}\}_p,
		\end{equation}
		where the Poisson bracket can be obtained mostly easily from the quantum commutators using the Dirac correspondence~\cite{dirac1981principles}
		\begin{equation}\label{eq:Dirac_corres}
			\{A,B\}_p \longleftrightarrow -i[\hat{A},\hat{B}].
		\end{equation}

		\item Substitute the following formula for the variables $\Phi_i$ in the equations of motion
		\begin{equation}\label{eq:HSfield}
			\Phi_i = \frac{1}{2}\sum_j \Gamma_{ij} L_j + \frac12 \xi_i,
		\end{equation}
		where $\xi_i$ is a Gaussian noise defined by
		\begin{equation}\label{eq:gauss_noise}
			\overline{\xi_i(t)}=0, \quad \overline{\xi_i(t) \bar{\xi}_j(t')} = 2\Gamma_{ij} \delta(t-t'),
		\end{equation}
        where bars here indicate  noise averages (not to be confused with the  same notation used for complex conjugation in Eq.~\eqref{eq:jm}). 
        
		\emph{We emphasize that the expression for $\Phi_i$ has to be substituted only after obtaining the equations of motion.} 
		
		\item The outcome of the above steps is
		\begin{multline}\label{eq:C_Langevin}
			\dv{}{t}\psi_{\alpha} = \{\psi_{\alpha},H\}_p - \frac{i}{2} \sum_i \{\psi_{\alpha},\bar{L}_i\}_p\Big( \sum_j \Gamma_{ij}L_j + \xi_i \Big) \\ - \frac{i}{2} \sum_i  \Big( \sum_j \bar{\Gamma}_{ij}\bar{L}_j + \bar{\xi}_i \Big)  \{L_i, \psi_{\alpha}\}_p,
		\end{multline}
		which is the semi-classical equation of motion, and the central result of this work.
		\item Sample initial conditions for classical variables according to the initial probability distribution function, to account for the quantum uncertainty in the initial state. For each initial condition, obtain a trajectory by evaluating stochastic dynamics according to Eq.~\eqref{eq:C_Langevin}.
		
		\item Obtain the expectation values of observables by taking their average over different trajectories and noise realizations.
		
	\end{enumerate}
	
	The initial distribution function can be obtained by taking the Wigner transformation of initial density matrix, as thoroughly discussed in Ref.~\cite{polkovnikov2010phase}. For spin-$1/2$ degrees of freedom, one can also use a discrete sampling of the initial state, which can yield improved results~\cite{wootters1987wigner,Schachenmayer2015}. We will discuss the sampling procedure for spins later in details, when we address spin systems as a special case in Sec.~\ref{sec:prot_spins}. 

     We emphasize that, the applicability of TWA is not limited to the calculation of single-point expectation values. In our approach, multi-point correlation functions of operators, possibly at different times, can be evaluated in terms of fully symmetrized (classical) correlators, together with (quantum) corrections which are obtained by measuring the response of the system to small jumps applied to the system. The latter step is similar to the calculation of response functions in classical stochastic dynamics~\cite{Kamenev_2023}. This procedure has been explained in details in Refs.~\cite{polkovnikov2010phase,WURTZ_ClusterTWA2018}.

\subsection{Connection to the quantum Langevin equation\label{sec:QLE}}
{We emphasize that the resulting protocol above is not merely an outcome of the field theory `black box'. Notably, using the Dirac correspondence in Eq.~\eqref{eq:Dirac_corres}, we can interpret
	Eq.~\eqref{eq:C_Langevin} as the semi-classical limit of QLE
\begin{multline}\label{eq:q_Langevin}
	\dv{}{t}\hat{\psi}_\alpha= i[\hat{H}, \hat{\psi}_\alpha] - \frac12 \sum_i \big[\hat{\psi}_\alpha, \hat{L}_i^{\dagger}\big]\Big( \sum_j \Gamma_{ij} \hat{L}_j+\hat{\xi}_i\Big) \\
	-\frac12 \sum_i\Big(\sum_j \bar{\Gamma}_{ij}\hat{L}_j^{\dagger}+\hat{\xi}_i^{\dagger}\Big)\big[\hat{L}_i, \hat{\psi}_\alpha\big],
\end{multline}
which describes the dynamics of quantum mechanical operators in the presence of dissipation, and is an alternative, but equivalent, representation of Lindblad master equation (Eq.~\eqref{eq:LindbladMaster}), with   quantum noise $\langle \hat{\xi}_i(t)\hat{\xi}_j^{\dagger}(t')\rangle=2\Gamma_{ij}\delta(t-t')$~\cite{gardiner1985handbook,walls2008quantum}. We recall that at the level of Eq.~\eqref{eq:q_Langevin} the noise, $\hat{\xi}_i(t)$, is an operator since it encapsulates the interaction between the quantum degrees of freedom of the system and of the environment (also operators in a microscopic description of system-bath coupling).

A crucial point to remember about the classical equation Eq.~\eqref{eq:C_Langevin} is that all variables are numbers, meaning they commute and should no longer be treated as operators. If, instead, we start with QLE~\eqref{eq:q_Langevin}, simplify the right-hand side using operator commutation relations, and take the classical limit afterwards, the resulting equations can be inconsistent with the semi-classical approximation~\cite{Huber_DDTWA} (cf. also with discussion in Sec.~\ref{sec:driven_spin}). We will discuss this issue in more details later in Sec.~\ref{sec:spins} for dissipative spins.

\section{Results for dissipative spins\label{sec:spins}}

In this section, we explore practical applications of our dissipative TWA to spin models. We begin in Sec.~\ref{sec:semiclassics} by specializing the protocol presented in Sec.~\ref{sec:semiclassics} to spins, together with remarks on the sampling of initial states. As a first example, Sec.~\ref{sec:driven_spin} examines a coherently driven spin subject to decay. We then analyze models with long-range interactions, including the Tavis-Cummings and central spin models, in Secs.~\ref{sec:TC}~and~\ref{sec:central}, respectively. In Sec.~\ref{sec:rydbergs}, we shift focus to short-range interactions, solving the dynamics of a driven Rydberg chain. Finally, Sec.~\ref{sec:corr_dissip} extends our approach beyond individual dissipation by considering an atomic chain with correlated emission.

\subsection{Protocol for spins\label{sec:prot_spins}}

Below, we mainly restate the rules of Sec.~\ref{sec:semiclassics} for spin degrees of freedom, which are also illustrated in Fig.~\ref{fig:schematics} for clarity. We will also briefly discuss the discrete sampling of initial conditions.
\begin{enumerate}
	\item Replace spin operators $\hat{\boldsymbol{\sigma}}_k=(\sigma^x_k,\sigma^y_k,\sigma^z_k)$ with the classical variables $\boldsymbol{s}_k=({s}_k^{x},{s}_k^{y},{s}_k^{z})$ and substitute them in the Hamiltonian and jump operators.
	\item The equation of motion for $s^\alpha_k$ follows from
	\begin{equation}
		\dv{}{t}s_k^{\alpha} = \{{s}_k^{\alpha},\tilde{H}\}_p,
	\end{equation}
	 with the effective classical Hamiltonian
	\begin{equation}
		\tilde{H}= H - i \sum_{i} \big(\bar{L}_i \Phi_i - \bar{\Phi}_i L_i \big).
	\end{equation}
	The Poisson's bracket of spin variables with another variable $O$ can be expressed as
	\begin{equation}
		 \{{s}_k^{\alpha},O\}_p=2\sum_{\beta,\gamma}\epsilon_{\alpha\beta\gamma}\frac{\partial O}{\partial s_k^{\beta}}s_k^{\gamma},
	\end{equation}
	where $\epsilon_{\alpha \beta \gamma}$ is the totally anti-symmetric tensor with $\epsilon_{xyz}=+1$.
	
	\item Substitute for $\Phi_i$ using Eqs.~\eqref{eq:HSfield}~and~\eqref{eq:gauss_noise} to get the classical Langevin equation for spins
	\begin{multline}\label{eq:C_Langevin_spins}
		\dv{}{t}s_k^{\alpha} = \{s_k^{\alpha},H\}_p - \frac{i}{2} \sum_i \{s_k^{\alpha},\bar{L}_i\}_p\Big( \sum_j \Gamma_{ij}L_j + \xi_i \Big) \\ - \frac{i}{2} \sum_i  \Big( \sum_j \bar{\Gamma}_{ij}\bar{L}_j + \bar{\xi}_i \Big)  \{L_i, s_k^{\alpha}\}_p,
	\end{multline}
	The final form of this equation is provided in Table~\ref{table} for some of the common dissipation channels. Since the equations of motion are obtained from an effective Hamiltonian, which is a function of spin variables, we have
	
	\begin{equation}\label{eq:spin_conserv}
		\dv{}{t} |\boldsymbol{s}_k|^2 = 4\sum_{\alpha \beta \gamma} \epsilon_{\alpha \beta \gamma} s_k^\alpha \frac{\partial \tilde{H}}{\partial s^\beta_k} s^\gamma_k = 0,
	\end{equation}
	due to the anti-symmetry property $\epsilon_{\alpha \beta \gamma}=-\epsilon_{\gamma \beta \alpha}$. Therefore, regardless of dissipation profile, the length of spins is always conserved for each trajectory, which is an essential condition for the consistency of TWA.
	\item In the next step, we sample initial conditions for $\boldsymbol{s}_{k,n}$ with $n=1,\dots,N_\mathrm{tr}$ to account for quantum uncertainty in the initial state, where $N_\mathrm{tr}$ is the number of sampled trajectories.
	\item For each trajectory, evaluate stochastic dynamics according to Eq.~\eqref{eq:C_Langevin}.
	\item Expectation values, and symmetric two-point functions are obtained from averaging over trajectories and noise realizations:
	\begin{equation}\label{eq:observables}
		\begin{aligned}
			\langle \hat\sigma _k^{\alpha}(t)\rangle&=\frac{1}{N_\mathrm{tr}}\sum_{n=1}^{N_\mathrm{tr}}\overline{s_{k,n}^{\alpha}(t) }\\
			 \frac12 \langle \{\hat\sigma _k^{\alpha}(t),\hat\sigma _l^{\beta}(t)\}\rangle &=\frac{1}{N_\mathrm{tr}}\sum_{n=1}^{N_\mathrm{tr}}\overline{s_{l,n}^{\alpha}(t) s_{k,n}^{\beta}(t)},
		\end{aligned}
	\end{equation}
	where overlines represent averaging over the noise, and $\{A,B\}=AB+BA$ is the anti-commutator. Fully symmetrized higher-order correlation functions can also be obtained in the same way as two-point functions. To access non-symmetric correlators with TWA, where the non-trivial commutation of operators is important, one has to take extra steps which are explained in Appendix~\ref{app:correlation}.
	
\end{enumerate}

\newcolumntype{L}[1]{>{\raggedright\arraybackslash}p{#1}}

\begin{table}[t!]
	\centering
	\begin{tabular}{|c|L{5.5cm}|} 
		\hline  \vspace{-6pt} & \vspace{-6pt} \\
		
		$\hat L_{i}$& \multicolumn{1}{c|}{{Equations of motion}}  
		\\ \vspace{-9pt} & \vspace{-9pt} \\ \hline  \vspace{-9pt} & \vspace{-9pt} \\

		$\sqrt{\gamma_\downarrow}\,\hat\sigma^-$         & 
		$ \begin{aligned}
			& \mathrm{d} s^x/\mathrm{d}t =\gamma_\downarrow s^x s^z/2+\xi_\downarrow^{x}s^z\\
			& \mathrm{d} s^y/\mathrm{d}t=\gamma_\downarrow s^y s^z/2+\xi_\downarrow^{y}s^z\\
			&\mathrm{d} s^z/\mathrm{d}t=-\gamma_\downarrow\, (s^x s^x+s^y s^y)/2 -(\xi_\downarrow^{x}s^x+\xi_\downarrow^{y}s^y)\\ &\overline{\xi_\downarrow^\alpha(t)\,\xi_\downarrow^\beta(t')}=\gamma_\downarrow\,\delta_{\alpha\beta}\,\delta(t-t')
		\end{aligned}$                      

		\\ \vspace{-11pt} & \vspace{-11pt} \\ \hline \vspace{-9pt} & \vspace{-9pt} \\

		$\sqrt{\gamma_\uparrow}\,\hat\sigma^+$         & 
		$ \begin{aligned}
			& \mathrm{d} s^x/\mathrm{d}t =-\gamma_\uparrow s^x s^z/2-\xi_\uparrow^{x}s^z\\
			& \mathrm{d} s^y/\mathrm{d}t =-\gamma_\uparrow s^y s^z/2+\xi_\uparrow^{y}s^z\\
			& \mathrm{d} s^z/\mathrm{d}t =\gamma_\uparrow\,(s^x s^x+s^y s^y)/2 +(\xi_\uparrow^{x}s^x-\xi_\uparrow^{y}s^y) \\ &\overline{\xi_\uparrow^\alpha(t)\,\xi_\uparrow^\beta(t')}=\gamma_\uparrow\, \delta_{\alpha\beta} \, \delta(t-t')
		\end{aligned}$            
		\\ \vspace{-9pt} & \vspace{-9pt} \\ \hline \vspace{-9pt} & \vspace{-9pt} \\

		$\sqrt{\kappa}\,\hat\sigma^z$           & 
		$ \begin{aligned}
			&\mathrm{d} s^x/\mathrm{d}t=+2\,\eta s^y\\
			&\mathrm{d} s^y/\mathrm{d}t=-2\,\eta s^x\\
			& \mathrm{d} s^z/\mathrm{d}t=0 \\ 			&\overline{\eta(t)\,\eta(t')}=\kappa\,\delta(t-t')
		\end{aligned}$    
		\\  \hline
	\end{tabular}
	\caption{Classical equations of motion for incoherent spin loss, spin pump and dephasing. Note that we have absorbed the rates into the jump operators such that $\Gamma_{ij}=\delta_{ij}$ in Eq.~\eqref{eq:LindbladMaster}. The real-valued noises originate from the real and imaginary parts of the complex noises introduced in Eq.~\eqref{eq:C_Langevin_spins}.}
	\label{table}
\end{table}

In practice, the sampling of initial state and the noise can be performed together, i. e., we can write
\begin{equation}
	\frac{1}{N_\mathrm{tr}}\sum_{n=1}^{N_\mathrm{tr}}\overline{s_{k,n}^{\alpha}(t) } = \frac{1}{N_\mathrm{tot}}\sum_{n'=1}^{N_\mathrm{tot}} s_{k,n'}^{\alpha}(t),
\end{equation}
where $N_\mathrm{tot}=N_\mathrm{tr}\times N_\mathrm{noise}$ and $N_\mathrm{noise}$ is the number of noise realizations. 

For spins, there exist two common sampling schemes, which are the continuous and discrete approaches (cf. Refs.~\cite{Fleischhauer_2022,mink2023collective} for a hybrid discrete-continuous scheme). The stating point of the continuous sampling consists of resorting to a bosonic representation of spins~\cite{polkovnikov2010phase,Rabl_phase_space_2021}. However, for spin coherent states, the following distribution has been found to be a good approximation~\cite{polkovnikov2010phase}
\begin{equation}\label{eq:continuous_Wigner_probability}
    P(\boldsymbol{s}) = \frac{e^{-s_\perp^2/4S} }{4\pi S}\,\delta(\boldsymbol{s}\cdot \boldsymbol{n} - 2S) ,
\end{equation}
where $\boldsymbol{n}$ is the unit-vector in the direction of the coherent state, $S$ is the spin size, and $s_\perp$ is the component of $\mathbf{s}$ orthogonal to $\boldsymbol{n}$. Note that in our convention, the average length of $\boldsymbol{s}$ is $2S$, such that in the case of spin-1/2 the classical vectors have unit length.

A discrete sampling, known as discrete TWA (DTWA)~\cite{Schachenmayer2015,wootters1987wigner,Sundar_TWA_DTWA}, is possible for spin-1/2 degrees of freedom. For instance, for the initial state with  $\sigma^z_k\ket{\downarrow}=-\ket{\downarrow}$, we sample the classical initial states according to the following discrete distribution
\begin{equation}\label{eq:discrete_Wigner_probability}
	W_0(s^x_k,s^y_k,s^z_k)= \begin{cases}
		0  & \quad(s^x_k,s^y_k,s^z_k)=(\pm 1,\pm1,+1),\\ 1/4 &\quad (s^x_k,s^y_k,s^z_k)=(\pm 1,\pm1,-1),
	\end{cases}
\end{equation}
which has been shown to yield improved results in certain cases, in comparison to the continuous sampling~\cite{Schachenmayer2015,Fleischhauer_2022}. Discrete sampling for $S>1/2$ spins can be also found, as reported in Ref.~\cite{zhu2019generalized}.

In the following sections, we use the discrete sampling, and solve Eq.~\eqref{eq:C_Langevin_spins} using an implicit numerical integration scheme and the Stratonovich regularization of stochastic dynamics~\cite{gardiner1985handbook,Kamenev_2023}.Using the Stratonovich regularization originates from the fact that the Markovian limit is an approximation to a non-Markovian bath whose correlation time is much shorter than the timescales of the system, as explained in Appendix~\ref{app:derivation}. We emphasize that \emph{our approach does not rely on any specific representation of spin operators, such as Schwinger bosons or spin coherent states, nor it is restricted to spin-1/2 degrees of freedom.} Likewise, our derivation in Appendix~\ref{app:derivation} is representation-independent, following directly from general field-theoretic arguments.

\subsection{Driven spin}\label{sec:driven_spin}

As a first example, we consider a coherently driven single spin. The exact solvability of this problem allows us to evaluate the accuracy of our semi-classical approximation and determine its range of applicability. The dynamics of the spin are governed by the following Hamiltonian
\begin{equation}
	\hat{H} = \Omega\, \hat{\sigma}^x,
\end{equation} 
with the Rabi frequency $\Omega$, together with incoherent loss ($\hat{L}^\downarrow=\hat{\sigma}^-$) with the rate $\gamma_\downarrow$. Using Eq.~\eqref{eq:C_Langevin_spins}, we obtain the following equations of motion for classical spin variables~$(s^x,s^y,s^z)$
\begin{align}
	\dv{}{t}s^x &= + \frac{\gamma_\downarrow}{2} s^x s^z + \xi_\downarrow^x s^z,\label{eq:EOM_single_x}\\
	\dv{}{t}s^y &= -2  \, \Omega \, s^z + \frac{\gamma_\downarrow}{2} s^y s^z + \xi_\downarrow^y s^z,\label{eq:EOM_single_y}\\
	\dv{}{t}s^z &= +2 \, \Omega \, s^y - \frac{\gamma_\downarrow}{2} \Big( (s^x)^2 + (s^y)^2\Big) - \xi_\downarrow^x s^x - \xi_\downarrow^y s^y,\label{eq:EOM_single_z}
\end{align}
where the noise fields satisfy
\begin{equation}
	\overline{\xi_\downarrow^\alpha(t)\, \xi_\downarrow^\beta(t')} = \gamma_\downarrow \delta_{\alpha \beta} \delta(t-t').
\end{equation}
Fig.~\ref{fig:driven_single_spin} compares the exact solution with semi-classical results obtained from Eqs.~\eqref{eq:EOM_single_x}-\eqref{eq:EOM_single_z} for different values of $\gamma_\downarrow$. We observe that TWA captures exact dynamics for weak to moderate losses ($\gamma_\downarrow \lesssim \Omega$), while it progressively deviates from the correct answer for larger decay rates. In the extreme limit $\Omega \to 0$, TWA captures early-time dynamics correctly, while deviating from the correct steady state at later times. The accuracy of the method also improves in the presence of incoherent pumping ($\hat{L}^\uparrow = \hat{\sigma}^+$), with the highest accuracy achieved when $\gamma_\uparrow = \gamma_\downarrow$.


\begin{figure}
	\centering
	\hspace{-15pt}\includegraphics[width=0.9\linewidth]{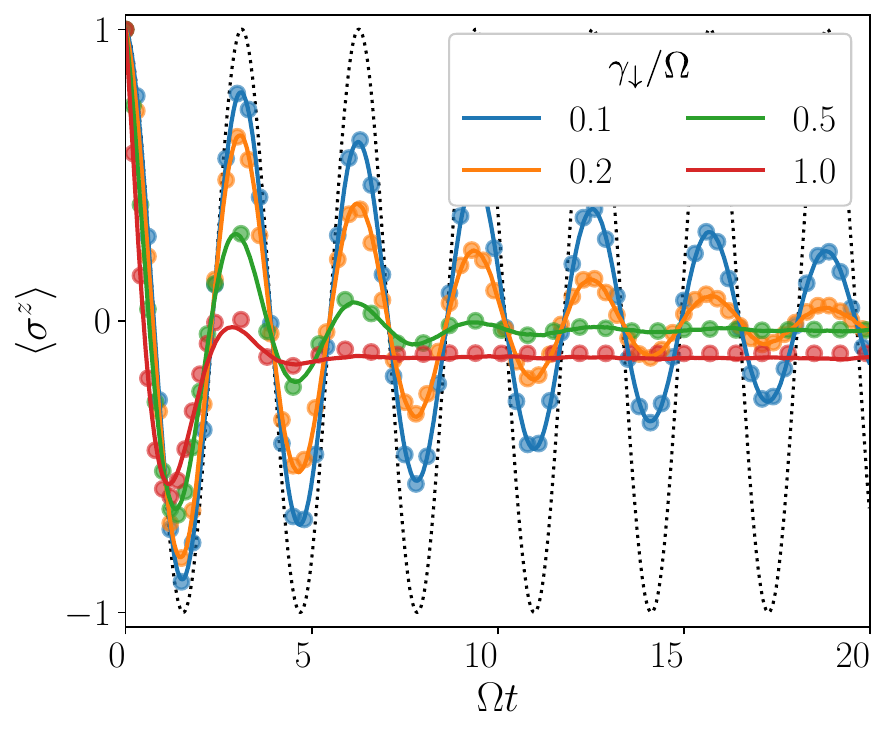}
	\caption{Evolution of a single spin under coherent driving and spontaneous decay. TWA (solid lines) shows strong agreement with the exact solution (circles) for weak to moderate loss rates ($\gamma_\downarrow \lesssim \Omega$). For $\gamma_\downarrow \gtrsim \Omega$, quantum fluctuations grow beyond the reach of our semi-classical approximation and cause deviations from the exact solution.  The dotted line corresponds to $\gamma_\downarrow/\Omega=0.1$ without including noise. The results have been obtained by averaging over $O(10^5)$ trajectories.
	}
	\label{fig:driven_single_spin}
\end{figure}

As we showed earlier in Eq.~\eqref{eq:spin_conserv}, our approach conserves the size of spins for each trajectory, since it takes the `proper' classical limit of QLE (Eq.~\eqref{eq:q_Langevin}). Therefore, it is instructive to consider also the `improper' classical limit of QLE for this problem, in order to juxtapose these two classical approximations. First, we substitute the spin and jump operators in Eq.~\eqref{eq:C_Langevin}, without simplifying the resulting equations:
\begin{align}\label{eq:Lang}
	\dv{}{t}\hat{\sigma}^x &=  + \frac12 \hat{\sigma}^z \big(\gamma_\downarrow \hat{\sigma}^- + \hat{\xi}_\downarrow\big) + \frac12 \big( \gamma_\downarrow \hat{\sigma}^+ + \hat{\xi}_\downarrow^\dagger \big) \hat{\sigma}^z,\\
	\dv{}{t}\hat{\sigma}^y &= -2 \, \Omega \, \hat{\sigma}^z + \frac{i}{2} \hat{\sigma}^z \big(\gamma_\downarrow \hat{\sigma}^- + \hat{\xi}_\downarrow\big) - \frac{i}{2} \big( \gamma_\downarrow \hat{\sigma}^+ + \hat{\xi}_\downarrow^\dagger \big) \hat{\sigma}^z,\\
	\dv{}{t}\hat{\sigma}^z &= +2 \, \Omega \, \hat{\sigma}^y -\hat{\sigma}^+ \big( \gamma_\downarrow \hat{\sigma}^- + \hat{\xi}_\downarrow \big)-  \big( \gamma_\downarrow \hat{\sigma}^+ + \hat{\xi}_\downarrow^\dagger \big) \hat{\sigma}^-.
\end{align}
If we now take the classical limits of these equations, we immediately get Eqs.~\eqref{eq:EOM_single_x}-\eqref{eq:EOM_single_z}, with $\xi_\downarrow^x$ and $\xi_\downarrow^y$ given by the real and imaginary parts of the complex-valued noise $\xi_\downarrow$. In contrast, if we proceed to simplify these equations by using the spin commutation relations for the non-noisy terms and then take the classical limit, we obtain
\begin{align}
	\dv{}{t}s^x &=  - \frac{\gamma_\downarrow}{2} s^x + \xi_\downarrow^x s^z,\label{eq:EOM_wrong_sx}\\
	\dv{}{t}s^y &= -2 \, \Omega \, s^z - \frac{\gamma_\downarrow}{2} s^y + \xi_\downarrow^y s^z,\\
	\dv{}{t}s^z &= +2 \, \Omega \, s^y-\gamma_\downarrow (1+s^z)  - \xi_\downarrow^x s^x - \xi_\downarrow^y s^y,\label{eq:EOM_wrong_sz}
\end{align}
which modifies the length of spin over time $\dv{}{t}s^2=-\gamma_{\downarrow} (s^2+(s^z)^2+2s^z)$, with $s^2=(s^x)^2+(s^y)^2+(s^z)^2$, leading to an incorrect limit at long times. To resolve this issue, previous works~\cite{Huber_DDTWA} introduced an ad-hoc modification to the noise terms, such that the spin-shrinkage rate becomes small. 

In contrast, noise naturally arises in our approach from expanding the Keldysh action of the system order by order in powers of quantum fields, as explained in Appendix~\ref{app:derivation}. The leading-order of the expansion yields the statistical sampling of the initial state, while noise emerges at the next-to-leading-order contribution. 
Consequently, the effect of noise becomes less significant when deterministic contributions dominate the dynamics. This occurs, for instance, when the system is strongly driven, ($\Omega \gtrsim \gamma_\downarrow$), as shown earlier, or in the collective-spin limit where $S$ is large, while rescaling $\gamma_{\downarrow}$ such that the limit $S \to \infty$ remains well-defined. In the latter case, dynamics become fully classical with mean-field equations being exact, and noise is negligible~\cite{Breuer_Petruccione}. As shown in Appendix~\ref{app:error_estimate}, the magnitude of quantum corrections to the classical value \emph{for this specific problem} is given by
\begin{equation}
    \mathrm{quantum\,\,corrections} \sim \frac{1}{2S}\sqrt{1-\qty(\frac{4\Omega}{\gamma_\downarrow})^2},
\end{equation}
which is suppressed for stronger drives and larger spins. Conversely, when quantum effects fully dominate, either when $S=1/2$ or $\Omega \ll \gamma_\downarrow$, higher-order corrections beyond noise must be included for accurate results. 
In the intermediate regime, where $S$ is small but $\Omega \gtrsim \gamma_\downarrow$, incorporating noise while discarding higher order effects is sufficient, whereas neglecting the noise leads to significant errors. This has been explicitly demonstrated in Fig.~\ref{fig:driven_single_spin}, where solving the equations without noise results in a dramatic deviation from the correct dynamics, even for weak dissipation. In Appendix~\ref{app:error_estimate} we also contrast the noise strength in the TWA and Heisenberg-Langevin approach to shed further light on the opposite limit, $ \gamma_\downarrow\gtrsim\Omega $.

We note that   some alternative approaches to TWA can accurately capture single-spin decay across all decay rates~\cite{Fleischhauer_2022,mink2023collective}. In our case, this would require to go beyond leading order in quantum fluctuations. However, this specific example of a single driven spin is presented for purely pedagogical reasons,  while for the practical purposes of solving driven-open many-body dynamics, both our approach and those in Refs.~\cite{Fleischhauer_2022,mink2023collective} achieve similar accuracy in comparable windows of parameters.

The following sections showcase how our approach can cover a broad variety of models, by applying the straightforward rules of Tab.~\ref{table}, with an efficiency that outperforms  competing techniques (such as cumulants dynamics).  These examples not only provide easy benchmarks   to   first-time users, but also illustrate in practice the efficiency of our universal DTWA program.

}

\subsection{Tavis-Cummings model}\label{sec:TC}

The first model we analyze is in the class of all-to-all interacting systems. 
As mentioned in the introduction, these models are particularly instrumental in systematically studying corrections on top of mean field dynamics, since they are usually equipped with a `large N' parameter~\cite{defenu,kirton2019introduction}.

 The Tavis-Cummings (TC) model is a prototypical model for lasing~\cite{Breuer_Petruccione,gardiner2004quantum} which finds applications in modern AMO research,
  in particular, to study ultra-narrow linewidth lasing in a bad cavity~\cite{bohnet2012steady,Holland_millihertz,PhysRevA.81.033847,Jager_bad_laser,hot_lasing} and dynamical phase transitions between non-radiative, lasing and superradiant lasing regimes~\cite{Larson_2017,Hannukainen_dissPT2018,Lasing_SR_crossover,kirton2018superradiant}.  
 TC consists of a single bosonic mode, representing photons, coupled to an ensemble of spin-$1/2$ degrees of freedom, modeling two-level atoms confined within an optical cavity. Its relevance in our narrative is   as a  first step in comparing the efficiency of TWA and CE in capturing the dynamics and steady states of driven-open systems.
 
 The TC Hamiltonian reads
\begin{equation}
	\hat{H} = \omega \, \hat{a}^\dagger \hat{a} + \frac{\epsilon}{2} \sum_{i=1}^N \hat{\sigma}^z_i + \frac{g}{\sqrt{N}} \sum_{i=1}^N \Big(\hat{a}^\dagger \hat{\sigma}^-_i  + \hat{a} \hat{\sigma}^+_i\Big).
\end{equation}
$\omega$ and $\epsilon$ are respectively the excitation energies of free photons and atoms, and $g$ is the light-matter coupling. Incoherent processes include photon loss ($\hat{L}^\kappa =\hat{a}$), atomic loss ($\hat{L}^{\downarrow}_i= \hat{\sigma}^-_i$), and atomic pumping ($\hat{L}^\uparrow_i = \hat{\sigma}^+_i$), with respective rates $\kappa$, $\gamma_\downarrow$, and $\gamma_\uparrow$. In the absence of dissipation, the TC model exhibits a $U(1)$ symmetry, defined by the transformation $(\hat{a},\hat{\sigma}^-_i) \to (e^{i\phi}\hat{a},e^{i\phi}\hat{\sigma}^-_i)$, which ensures the conservation of the total number of excitations, $\hat{n}_\mathrm{tot}=\hat{a}^\dagger \hat{a} +  \sum_i \hat{\sigma}^z_i/2$. Additionally, the model possesses permutation symmetry under spin exchange. These symmetries render the model integrable, allowing its solution by separately analyzing subspaces of the Hilbert space with fixed excitation number and total angular momentum.

\begin{figure}[!t]
	\centering
	\hspace{-15pt}\includegraphics[width=0.9\linewidth]{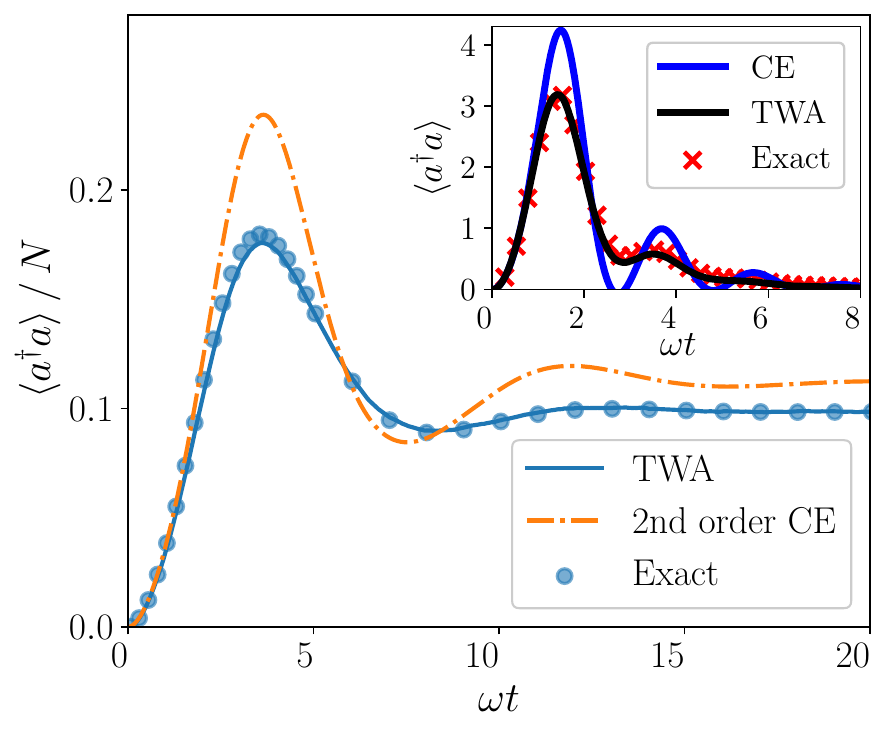}
	\caption{The evolution of the normalized photon population in the dissipative Tavis-Cummings model with \( N=15 \), starting with the photonic vacuum with fully inverted atoms. TWA demonstrates excellent agreement with the exact solution, whereas the second-order CE becomes inaccurate beyond short times. The parameters used are \( \omega = \epsilon = \kappa \), \( g/\omega = 9/10 \), \( \gamma_\downarrow/\omega = 1/8 \), and \( \gamma_\uparrow/\omega = 3/8 \). Inset: Dynamics of photon population without incoherent pump ($\gamma_\uparrow=0$) for $N=12$ spins, showing good agreement between TWA and the exact result. The other parameters are \( \omega = \kappa = 2\epsilon \), \( g/\omega \approx 1.7 \), and \( \gamma_\downarrow/\omega = 0.25 \). TWA results are obtained from \( O(10^5) \) trajectories.}
	\label{fig:TC_comp}
\end{figure}

However, dissipation typically breaks these symmetries. Photon loss only breaks $U(1)$ symmetry while preserving permutation symmetry, still making exact solutions feasible for sufficiently large system sizes. In contrast, individual atomic decay breaks both symmetries, significantly complicating the solution. Although weak permutation symmetry can still be exploited to reduce computational costs for steady-state calculations~\cite{kirton2017suppressing,PhysRevResearch.5.033148}, exact time evolution remains accessible only for small systems. Common approximation methods, such as MF theory and cumulant expansion (CE), often yield inconsistent results, which also depend on the order of CE~\cite{PhysRevResearch.5.033148}. In some cases, these approximations approach the correct result in the $N\to \infty$ limit, but they typically exhibit slow convergence. That is, even for large but finite system sizes, deviations from exact solutions persist over a broad parameter range.

To solve the problem with TWA, we substitute photon and spin variables into Eq.~\eqref{eq:C_Langevin} to get the following equations of motion
\begin{align}
	\dv{}{t} a &= -i \omega a - \frac{ig}{\sqrt{N}}\sum_i s^-_i - \frac{\kappa}{2} a -\frac12 \xi_\kappa,\\
	\dv{}{t}s^x_i &= -\epsilon s^y_i - \frac{2g}{\sqrt{N}} s^z_i \,\mathrm{Im}\,a+\frac{\gamma_\downarrow-\gamma_\uparrow}{2} s^x_i s^z_i  + \big(\xi_{\downarrow,i}^x - \xi_{\uparrow,i}^x\big)s^z_i,\\
	\dv{}{t}s^y_i &= +\epsilon s^x_i - \frac{2g}{\sqrt{N}} s^z_i \,\mathrm{Re}\,a+\frac{\gamma_\downarrow-\gamma_\uparrow}{2} s^y_i s^z_i  + \big(\xi_{\downarrow,i}^y + \xi_{\uparrow,i}^y\big)s^z_i,\\
	\dv{}{t}s^z_i &= + \frac{4g}{\sqrt{N}} \,\mathrm{Im}\,\big( a \,s^+_i \big) -\frac{\gamma_\downarrow-\gamma_\uparrow}{2} \Big( (s^x_i)^2 + (s^y_i)^2 \Big) \nonumber\\ & \qquad \qquad \qquad + \big(\xi^x_{\uparrow,i}-\xi^x_{\downarrow,i}\big)s^x_i  - \big(\xi^y_{\uparrow,i}+\xi^y_{\downarrow,i}\big)s^y_i ,
\end{align}
where the noises are given by
\begin{align}
	\overline{ \xi_\kappa(t)\bar{\xi}_\kappa(t')} &= 2\kappa \, \delta(t-t'),\\
	\overline{ \xi_{\downarrow,i}^\alpha(t) \xi_{\downarrow,j}^\beta (t')} & = \gamma_\downarrow \, \delta_{ij} \,\delta_{\alpha \beta} \,\delta(t-t'),\label{eq:gammaDown_noise}\\
	\overline{ \xi_{\uparrow,i}^\alpha(t) \xi_{\uparrow,j}^\beta(t')} &=\gamma_\uparrow \, \delta_{ij} \, \delta_{\alpha \beta} \,\delta(t-t').\label{eq:gammaUp_noise}
\end{align}
In Fig.~\eqref{fig:TC_comp}, we have shown TWA results for the dynamics of photon population, which can be obtained from the trajectory-average via
\begin{equation}
	\langle \hat{a}^\dagger\hat{a} \rangle =\frac{1}{N_\mathrm{tr}} \sum_{n=1}^{N_\mathrm{tr}} \overline{|a_{n}|^2} - \frac12.
\end{equation}
Comparison with the exact solution shows that TWA accurately captures both the transient dynamics and the steady state for all system sizes, with a relative error of $O(10^{-2})$ which can be reduced by including more trajectories. We also see that CE is accurate only for short-time dynamics, predicting an incorrect steady-state value. It has been shown~\cite{PhysRevResearch.5.033148} that, for the same model, higher order CE either are similar and converge slowly, or become unstable for sufficiently large values of $N$.

\subsection{Central spin model}\label{sec:central}

For the purpose of refining our comparison between TWA and CE, we solve the central spin (CS) model, describing a single spin interacting collectively with an ensemble of $N$ satellite spins, which is a prototypical model of single spins or dilute spin ensembles interacting with many-body environments. Examples include nuclear magnetic resonance and nitrogen-vacancy centers in diamond~\cite{Childress_NV2006,Hall_centralspin2014}. In the simplest case, the CS Hamiltonian is given by
\begin{equation}
	\hat{H} = \frac{\omega}{2}\hat{\tau}^z + \frac{\epsilon}{2} \sum_{i=1}^N \hat{\sigma}^z_i + \frac{g}{\sqrt{N}} \sum_{i=1}^N \Big(\hat{\tau}^+ \hat{\sigma}^-_i  + \hat{\tau}^- \hat{\sigma}^+_i\Big),
\end{equation}
where $\hat{\tau}^\alpha$ are Pauli operators of the central spin.  In the following, we consider  central spin loss ($\hat{L}^{\kappa}=\hat{\tau}^-$) together with satellite spin loss ($\hat{L}^{\downarrow}_i=\hat{\sigma}^-_i$) and pumping ($\hat{L}^\uparrow_i =  \hat{\sigma}^+_i$) with respective rates $\kappa$, $\gamma_\downarrow$, and $\gamma_\uparrow$. The CS model can be obtained from the TC model by replacing $ \hat{a} \to \hat{\tau}^-$, and has the same symmetries. Due to the smaller local Hilbert space of the central spin versus the photon mode, we generally expect quantum fluctuations to be more important in the former case. For the TC model, as we go to larger values of $N$, a Gaussian state progressively becomes a better approximation for photons, improving the performance of CE. In contrast, no similar improvement in CE performance is expected when applied to the CS model (unless the central spin is large). This is the key reason why the comparison between the dynamics of the TC and CS model is particularly instructive in assessing the validity of TWA versus CE.

\begin{figure}[!t]
	\centering
	\hspace{-15pt}\includegraphics[width=0.9\linewidth]{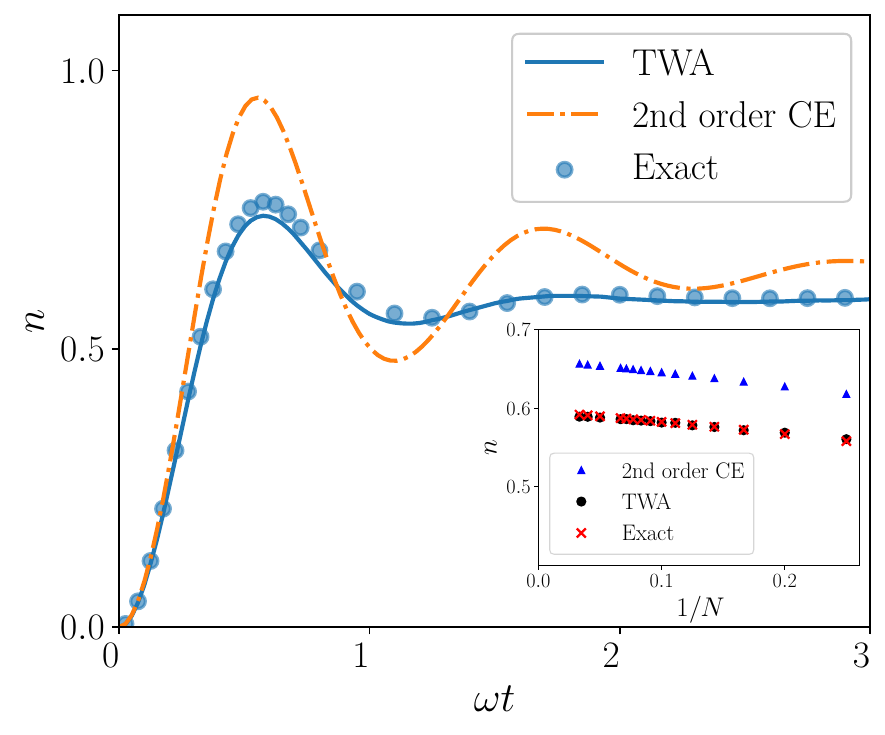}
	\caption{The population of the central site in the central spin model with \( N=30 \) as a function of time, starting from the central spin in its ground state and fully inverted satellite spins. TWA accurately captures the dynamics, whereas the second-order CE quickly deviates from the correct behavior. The inset shows the steady-state value for varying system sizes, where TWA remains accurate while CE fails across all system sizes. The parameters used are \( \omega = \epsilon = \kappa \), \( g/\omega = 3 \), \( \gamma_\downarrow/\omega = 1/2 \), and \( \gamma_\uparrow/\omega = 3/2 \). TWA results are obtained from \( O(10^5) \) trajectories.
	}
	\label{fig:CS_comp}
\end{figure}

We evaluate  the dynamics of the CS model, after initializing the central and satellite spins in their ground and excited states, respectively. The equations of motion are given by
\begin{align}
	\dv{}{t} \tau^x &= - \omega \tau^y + \frac{g}{\sqrt{N}} \tau^z \sum_i s^y_i + \Big(\frac{\kappa}{2}\tau^x   +  \xi^x_\kappa \Big) \,\tau^z,\\
	\dv{}{t} \tau^y &= + \omega \tau^x - \frac{g}{\sqrt{N}} \tau^z \sum_i s^x_i +\Big( \frac{\kappa}{2}\tau^y   +  \xi^y_\kappa \Big)\, \tau^z,\\
	\dv{}{t} \tau^z &=  + \frac{g}{\sqrt{N}}\Big( \tau^y \sum_i s^x_i  -  \tau^x \sum_i s^y_i \Big) - \frac{\kappa}{2}\Big((\tau^x)^2+(\tau^y)^2\Big) \nonumber\\ & \qquad \qquad \qquad \qquad \qquad \qquad \quad - \big(   \xi^x_\kappa \tau^x + \xi^y_\kappa \tau^y \big),
\end{align}
\begin{align}
	\dv{}{t}s^x_i &= -\epsilon s^y_i + \frac{g}{\sqrt{N}} \tau^y s^z_i +\frac{\gamma_\downarrow-\gamma_\uparrow}{2} s^x_i s^z_i  + \big(\xi_{\downarrow,i}^x - \xi_{\uparrow,i}^x\big)s^z_i,\\
	\dv{}{t}s^y_i &= +\epsilon s^x_i - \frac{g}{\sqrt{N}} \tau^x s^z_i +\frac{\gamma_\downarrow-\gamma_\uparrow}{2} s^y_i s^z_i  + \big(\xi_{\downarrow,i}^y + \xi_{\uparrow,i}^y\big)s^z_i,\\
	\dv{}{t}s^z_i &= + \frac{g}{\sqrt{N}}  \big(\tau^x s^y_i - \tau^y s^x_i\big) -\frac{\gamma_\downarrow-\gamma_\uparrow}{2} \Big( (s^x_i)^2 + (s^y_i)^2 \Big) \nonumber\\ & \qquad \qquad \quad \qquad + \big(\xi^x_{\uparrow,i}-\xi^x_{\downarrow,i}\big)s^x_i - \big(\xi^y_{\uparrow,i}+\xi^y_{\downarrow,i}\big)s^y_i .
\end{align}
 The noise variances are specified by
\begin{equation}
	\overline{\xi^\alpha_\kappa(t)\xi^\beta_\kappa(t')} = \kappa \, \delta_{\alpha\beta}\, \delta(t-t'),
\end{equation}
together with Eqs.~\eqref{eq:gammaDown_noise}~and~\eqref{eq:gammaUp_noise}. The TWA results for the central spin population, $n = (1+\langle \hat{\tau}^z\rangle)/2$, are shown in Fig.~\ref{fig:CS_comp}, demonstrating strong agreement with exact solutions for both transient dynamics and the steady state, with a relative error of $O(10^{-2})$. In contrast, second-order CE captures only the early-time dynamics but fails to predict the correct steady-state behavior for any system size, a limitation which persists even at higher orders of CE~\cite{PhysRevResearch.5.033148}.

The last two examples illustrate the improved reliability of TWA compared to CE. While for CE, going beyond the second order quickly increases the complexity of equations without guaranteed improvement~\cite{PhysRevResearch.5.033148,pavskauskas2012equilibration}, the equations for TWA are the simple MF equations supplemented by noise, and grant access to higher-point correlation functions without extra effort. Incorporating inhomogeneities, which naturally arise in realistic scenarios, further highlights the advantage of TWA over CE. In the CS model, such inhomogeneities can manifest in both the satellite spin splittings ($\epsilon \to \epsilon_i$), coupling strengths ($g\to g_i$) or dissipation rates ($\gamma_{\uparrow\downarrow}\to \gamma_{\uparrow\downarrow,i}$). While TWA seamlessly accommodates these variations, CE faces a significant increase in computational complexity, with the number of equations growing from 4 to approximately $N(N+3)/2$ at second order, and even more for higher-order expansions.

\subsection{Rydberg chain\label{sec:rydbergs}}

So far, we have considered models with collective interactions, where only local dissipation disrupts their fully collective nature. Next, we examine two prominent examples of systems with strongly non-collective behavior, whose physics crucially depends on the probed time and length scales, and which therefore qualify as genuine strongly correlated many-particle systems.  The next section is devoted to dissipation with non-trivial spatial structure.
Here, we consider a driven-dissipative chain of spins with short-range Ising-like interactions, describing the physics of Rydberg atomic arrays, which is one the most versatile platforms for quantum simulation and quantum computing nowadays~\cite{browaeys2020many}.

We consider the following Ising Hamiltonian
\begin{equation}\label{eq:Rydbergs}
	\hat{H}=\Omega \sum_i \hat{\sigma}_i^x+\frac{J}{4} \sum_{i} (1+\hat{\sigma}_i^z)(1+\hat{\sigma}_{i+1}^z),
\end{equation}
where $J$ is the strength of Rydberg interactions and $\Omega$ is the Rabi frequency. We consider local dephasing ($\hat{L}_i^\kappa=\hat{\sigma}_i^z,$) and spin loss ( $\hat{L}_i^\downarrow=\hat{\sigma}_i^{-}$) with rates $\kappa$ and $\gamma_\downarrow$, respectively. Furthermore, we work with periodic boundary conditions. The evolution simulated with TWA shows good agreement with the exact solution both for transient dynamics and the steady state, with the exception of regimes where   spontaneous decay is     stronger than the typical energy scales of coherent dynamics.

\begin{figure}
	\centering
	\hspace{-20pt}\includegraphics[width=0.9\linewidth]{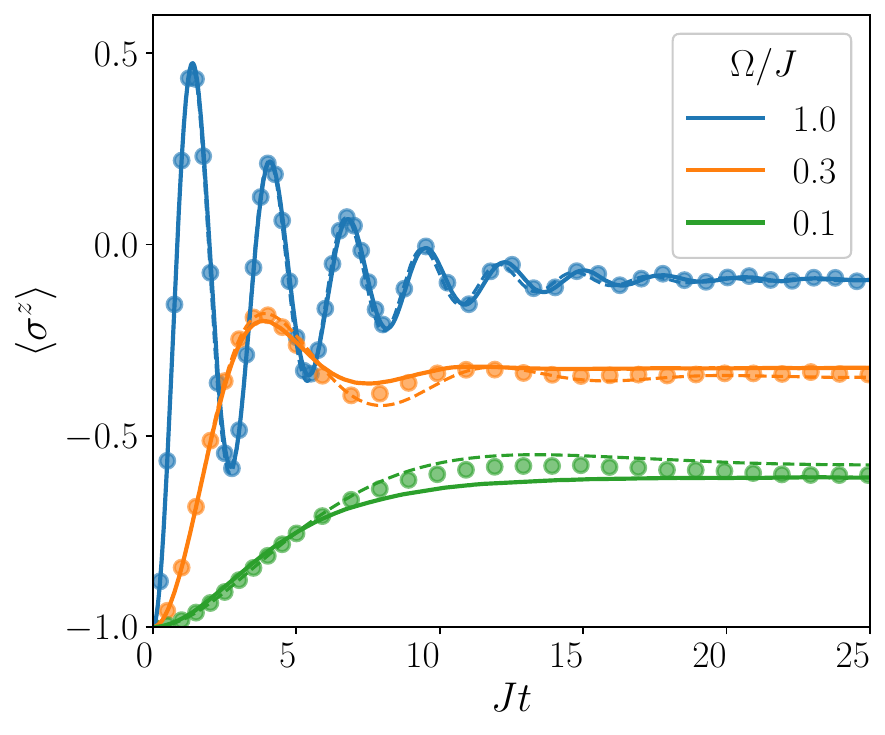}
	\caption{Dynamics of a Rydberg chain with size $N=10$, after being initialized in the atomic groundstate ($\ket{\downarrow}$).  Both TWA (solid lines) and CE (dashes) are in agreement with the exact results (circles). The parameters are $\gamma_\downarrow/J=\kappa/J=0.1$. The exact dynamics and TWA  have been respectively obtained by averaging over $O(10^3)$ quantum and $O(10^4)$ classical trajectories.}
	\label{fig:Rydbergs}
\end{figure}

Applying our formalism  to Eq.~\eqref{eq:Rydbergs} yields the following set of equations for classical variables 
\begin{align}
		\dv{}{t} s_i^x &= - \frac{J}{2} \Big( 2+ s^z_{i-1} + s^z_{i+1} \Big)\, s_i^y+\Big(\frac{\gamma_\downarrow}{2}s_i^x +\xi_{\downarrow,i}^{x}\Big) \,s_i^z+2\eta_i s_i^y,\\
			\dv{}{t} s_i^y &= - 2 \Omega s^z_i + \frac{J}{2} \Big( 2+ s^z_{i-1} + s^z_{i+1} \Big)\, s_i^x+\Big(\frac{\gamma_\downarrow}{2}s_i^y +\xi_{\downarrow,i}^{y} \Big) \, s_i^z-2\eta_i s_i^x,\\
		\dv{}{t} s_i^z &= +2\Omega s_i^y - \Big(  \frac{\gamma_\downarrow}{2} s^x_i + \xi_{\downarrow,i}^x \Big) \, s^x_i - \Big(  \frac{\gamma_\downarrow}{2} s^y_i + \xi_{\downarrow,i}^y \Big) \, s^y_i,
\end{align}
with dephasing ($\eta_i$) and spin loss ($\xi^\alpha_{\downarrow,i}$) noise variances given according to Table.~\ref{table}.

Figure~\ref{fig:Rydbergs} presents the dynamics of $\langle \hat{\sigma}^z \rangle$ for various driving strengths, as obtained from TWA, second-order CE, and the exact solution computed via the stochastic unraveling of quantum trajectories, using the  QuTiP  package of Ref.~\cite{Qutip}. For $\Omega/J \gtrsim 0.1$, TWA closely matches the exact dynamics, whereas for weaker drives (not shown), its accuracy progressively decreases (cf. Sec.~\ref{sec:driven_spin}), although it  improves upon  introducing spin dephasing.

CE performs similarly to TWA, and fails when interactions become very strong.
Notably, for this problem, implementing TWA is simpler than CE, which requires tracking all first-order ($\langle \hat{\sigma}^\alpha_i\rangle$) and second-order ($\langle \hat{\sigma}^\alpha_i \hat{\sigma}^\beta_j \rangle$) cumulants to obtain a closed system of equations. To handle the large number of the resulting equations, we used the QuantumCumulants.jl package~\cite{plankensteiner2022quantumcumulants} to obtain CE results. However, extending CE to higher orders proved considerably more computationally expensive than both TWA and the exact solution, even for a system of size $N=10$. The challenge stems in part from the fact that the current CE implementation in Ref.~\cite{plankensteiner2022quantumcumulants} cannot exploit the translation symmetry of the problem and calculates many redundant cumulants. We also note that translation symmetry is present only for homogeneous systems and under periodic boundary conditions, which do not apply in many situations. Without leveraging the translation symmetry, the symbolic evaluation of second-order CE for $N\gtrsim 20$ already exceeds the memory capacity of a personal computer. In contrast, TWA allows seamless simulation of dynamics for hundreds of spins on a desktop and several thousand on a supercomputer.

\subsection{Correlated decay \label{sec:corr_dissip}}

 As the final example of this work, we consider decay processes with a non-diagonal dissipation matrix in the atomic basis:
\begin{equation}
	\Gamma_{ij} \neq \gamma_i \,\delta_{ij}.
\end{equation}
A special case corresponds to fully collective decay~\cite{Haroche_SR,holzinger2025exactanalyticalsolutiondicke}, where $\Gamma_{ij}$ is uniform, and the analysis is simplified by recasting dissipation in terms of a collective jump operator $\hat{L}_\mathrm{col}=\sum_i \hat{\sigma}^-_i$.  This leads to the Dicke superradiance of  atomic ensembles, whose key features can be captured using simple approaches such as MF. Here, we are interested in regimes away from the limits of independent and fully collective decay. While this regime can be realized in different contexts~\cite{Asenjo_subrad2018,iemini2018boundary,henriet2019critical,Kushal1,Kushal2,marino_universality2022,bohnet2012steady,Liu_PRL2020,ferioli2024emergence,Jager_bad_laser}, here we focus on the particular example of correlated emission in sub-wavelength arrays of atoms~\cite{Asenjo_PRL,Asenjo_subrad2018,Yelin_cumulants,near_Dicke_SR,bach2024emergence,cascadedSR,agarwal2024directional,goncalves2024driven}.

\begin{figure}
	\centering
	\hspace{-20pt}\includegraphics[width=0.9\linewidth]{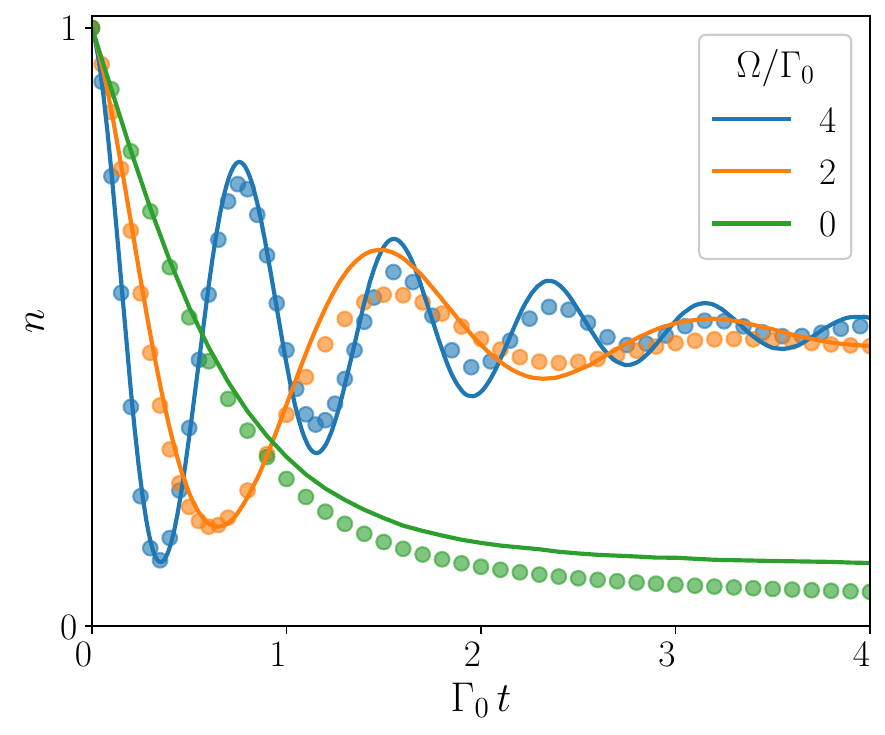}
	\caption{Dynamics of the average excitation number of a chain of $N=8$ emitters, subject to correlated decay and Rabi driving. Solid lines and circles  respectively correspond to TWA and the exact solution. The lattice spacing is $a=\lambda/5$, and the averages are taken over $O(10^4)$ trajectories.}
	\label{fig:driven_cor_em}
\end{figure}

The dynamics of a coherently driven atomic array placed in the electromagnetic (EM) vacuum can be expressed using the Lindblad master equation~\cite{Asenjo_subrad2018}
\begin{equation}
	\dv{}{t}\hat{\rho} = -i \,[\hat{H},\hat{\rho}]  + \sum_{ij=1}^N \Gamma_{ji}\, \Big(\hat \sigma_{i}^{-}\hat \rho\hat \sigma_{j}^{+}-\frac12\{\hat\sigma_{j}^{+}\hat \sigma_{i}^{-},\hat\rho\}\Big).
\end{equation}
The Hamiltonian reads
\begin{equation}
	\hat{H}=\omega_{z}\sum_{i=1}^N\hat\sigma_{i}^{z} + \Omega \sum_{i=1}^N \hat{\sigma}^x_i+\sum_{ij=1}^NJ_{ij}\,\hat\sigma_{i}^{+}\hat\sigma_{j}^{-},
\end{equation}
where $J_{ij}$ describes dipolar interaction between atoms, $\omega_z$ is the effective detuning which also includes the Lamb's shift due to coupling to EM modes, and $\Omega$ is the Rabi frequency. The interaction and dissipation matrices can be obtained from (in units where $\hbar\equiv 1$)
\begin{align}
	J_{ij} &= -\omega^2\, \bar{\boldsymbol{p}}\cdot \mathrm{Re}\, \boldsymbol{G}\big(\boldsymbol{r_i}-\boldsymbol{r_j},\omega\big)\cdot \boldsymbol{p},\\
	\Gamma_{ij} &= +2\omega^2\, \bar{\boldsymbol{p}}\cdot \mathrm{Im}\, \boldsymbol{G}\big(\boldsymbol{r_i}-\boldsymbol{r_j},\omega\big)\cdot \boldsymbol{p},
\end{align}
where $\boldsymbol{p}$ is the atomic dipole, and $\boldsymbol{G}$ is the Green's function of EM modes~\cite{jack}. In the vacuum we have
\begin{multline}
		\mathbf{G}_0(\mathbf{r},\omega)=\dfrac{\mu_0 \,e^{ikr}}{4\pi k^2 r^3}\Big[  (k^2r^2+ikr-1) \mathds{1}+(3-3ikr-k^2r^2)\frac{\mathbf{r}\otimes\mathbf{r}}{r^2}    \Big],
\end{multline}
 where $\omega$ is the optical transition frequency, $\mu_0$ is   the vacuum permeability, and $k=\omega/c$ is the momentum of emitted photons.
The physics of such a system crucially depends on the timescales and length-scales under consideration. In particular, the ratio of the inter-atomic distance $a$ to the wavelength of emitted light $\lambda \equiv 2\pi/k$ tunes the degree of collective behavior in the system. For $a\lesssim \lambda$, subsystems smaller than $\lambda$ can emit superradiantly, if initialized in a sufficiently excited state, while a fraction of the energy may remain trapped in the system for extended durations due to the existence of subradiant (evanescent) modes, depending on the system's geometry~\cite{Asenjo_subrad2018,henriet2019critical,Asenjo_PRL}. In the limit $a\gtrsim \lambda$, correlations become unimportant and atoms emit independently with the   rate
\begin{equation}
	\Gamma_0 = \lim_{j\to i} \Gamma_{ij} = \frac{\mu_0 \omega^3 |\boldsymbol{p}|^2}{3\pi c}.
\end{equation}

We simulate the dynamics of a one-dimensional array using our approach. The equations of motion for classical spin variables read (Sec~\ref{sec:prot_spins})
\begin{equation}\label{eq:correm}
	\begin{aligned}
		\dv{}{t} s_i^x&=-2\omega_z s_{i}^{y} + s_{i}^{z}\sum_{j}J_{ij}s_{j}^{y}
		+\frac{1}{2}s_{i}^{z}\sum_{j}\Gamma_{ij}s_{j}^{x}+\xi_{i}^{x}s_{i}^{z},\\
		\dv{}{t} s_i^y&=+2\omega_z s_{i}^{x} - 2 \Omega s_i^z  -s_{i}^{z}\sum_{j}J_{ij}s_{j}^{x}
		+\frac{1}{2}s_{i}^{z}\sum_{j}\Gamma_{ij}s_{j}^{y}+\xi_{i}^{y}s_{i}^{z},
		\\
		\dv{}{t} s_i^z&=+ 2 \Omega s_i^y +  \sum_{j}J_{ij}\,\Big( s^y_i s_{j}^{x}-s_i^{x}s_{j}^{y}\Big)-\frac12 \sum_{j}\Gamma_{ij}\,\Big(s_{i}^{x}s_{j}^{x}+s_{i}^{y}s_{j}^{y}\Big)\\ & \qquad \qquad \qquad\ \qquad \qquad\qquad \quad-\left(\xi_{i}^{x}s_{i}^{x}+\xi_{i}^{y}s_{i}^{y}\right),
	\end{aligned}
\end{equation}
with non-local noise variances
\begin{equation}
	\overline{ \xi^\alpha_i(t)\, \xi^\beta_j(t')} = \Gamma_{ij}\, \delta_{\alpha\beta}\,\delta(t-t').
\end{equation} 
Numerically, non-local noises can be sampled by diagonalizing the dissipation matrix
\begin{equation}\label{eq:gamma_diag}
	\hat{\Gamma}\cdot \boldsymbol{v}^j = \gamma_j \,\boldsymbol{v}^j,\quad 1 \le j \le N,
\end{equation}
with eigenvectors $\boldsymbol{v}^j$ and their corresponding eigenvalues $\gamma_i\ge 0$. We then define the collective noise variables according to
\begin{equation}
	\overline{ \eta^\alpha_i(t)\, \eta^\beta_j(t')} = \gamma_{i} \,\delta_{ij} \, \delta_{\alpha \beta}\, \delta(t-t'),
\end{equation}
such that the noise tensor in the site-basis can be obtained from the linear superposition of the collective noises via
\begin{equation}
	\xi_i^\alpha = \sum_{j=1}^N v^{\,j}_i \,\eta^\alpha_j.
\end{equation}

\begin{figure}
    \centering
    \hspace{-20pt}\includegraphics[width=0.9\linewidth]{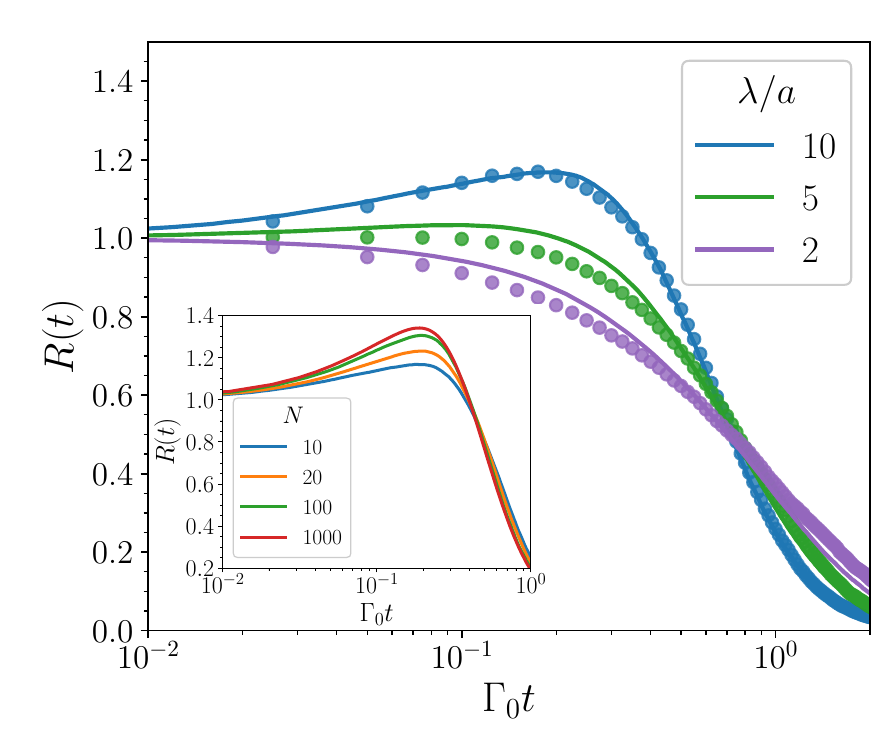}
    \caption{Emission rate of an initially inverted atomic chain with size $N=10$ as a function of time and for different values of lattice spacing. TWA (solid lines) captures the superradiant burst and matches   exact numerics (circles). Inset: TWA results for the emission rate of a fully inverted atomic chain with $\lambda/a=10$ as a function of time and for different system sizes.}
    \label{fig:burst}
\end{figure}

In Fig.~\ref{fig:driven_cor_em}, we compare the predictions of TWA for a sub-wavelength chain of driven atoms ($a=\lambda/5$) to exact results. We observe that for sufficiently strong driving relative to $\Gamma_0$, TWA accurately captures the system’s dynamics. For weak driving, it becomes less reliable and, as in the single-spin case discussed in Sec.~\ref{sec:driven_spin}, predicts an incorrect steady-state population. However, we remark that the accuracy of TWA for this problem cannot be assessed by a single ratio such as $\Omega/\Gamma_0$, as it depends crucially on the spatial structure of the eigenvectors in Eq.~\eqref{eq:gamma_diag}. Importantly, the collective decay rates in Eq.~\eqref{eq:gamma_diag} span a broad range of energy scales~\cite{henriet2019critical,Asenjo_PRL}, with $\Gamma_0$ representing only their average, $\Gamma_0 = \sum_i \gamma_i / N$. Notably, in the limit $a\ll \lambda$, where the largest eigenvalue of $\Gamma$ corresponds to the most superradiant mode, TWA correctly captures the superradiant burst even in the absence of a drive ($\Omega=0$).  To illustrate that, we plot the normalized emission rate
\begin{equation}\label{eq:rate}
\begin{aligned}
    R(t)&\equiv-\frac{1}{N\Gamma_0}\dv{}{t} n=\frac{1}{N\Gamma_0}\sum_{ij}\Gamma_{ij}\langle \sigma_i^+\sigma_j^-\rangle=\\
&=\frac{1}{N\Gamma_0}\sum_{ij}\left[\frac{1}{N_{\text{tot}}}\sum_{n=1}^{N_{\text{tot}}}\Gamma_{ij}s_{i,n}^+s_{j,n}^-+\frac{\Gamma_0}{2N_{\text{tot}}}\sum_{n=1}^{N_{\text{tot}}}s_{i,n}^z\delta_{ij}
    \right]
\end{aligned}
\end{equation}
for different values of $a$ and $N=10$ in Fig.~\ref{fig:burst}. In the third step of ~\eqref{eq:rate} we transited from operators to classical variables. For all simulations, TWA captures essential features of the dynamics, and for small atomic separations, $a=\lambda/10$, it matches quantitatively the exact numerical dynamics of   $R(t)$. In the inset, we plot dynamics up to $N=1000$ atoms, but larger system sizes are also at reach with  our approach, by running numerics on a supercomputer. Importantly, TWA requires solution of $3N$ stochastic differential equations, and the operational capabilities  of the method are set mostly by the accessible memory resources. On the other hand, third-order CE   employed  recently to simulate  correlated emission~\cite{Yelin_cumulants}, would saturate faster, as the number of equations to be solved scales like $N^3$. The intricacy of higher order CE (see Appendix of ~\cite{Yelin_cumulants}) compared to Eqs.~\eqref{eq:correm} would also pose a challenge  to apply cumulants to other Lindbladians with non-trivial spatial structure~\cite{Kushal1}. 
Nevertheless, cumulants succeed to capture the  late time dynamics approaching the vacuum state~\cite{Yelin_cumulants} (`sub-radiance'), where TWA fails. This is in line with our earlier remarks on including higher order corrections to access the small pumping limit, see Sec.~\ref{sec:driven_spin}.

\section{Conclusions and outlook\label{sec:conclusions}}

In recent years, there has been a growing body of work on solving driven dissipative spin models using semi-classical methods, and increasing evidence of their capability to fit AMO experiments~\cite{Rabl_phase_space_2021,mink2023collective,Fleischhauer_2022,Huber_DDTWA,PhysRevResearchRey,Singh_OSDTWA2022,cascadedSR,windt2024effects,kamar2023hybrid,kessler2021observation,Kessler2020From,Kessler2019Emergent,NagaoKazuma_bosons,robicheaux2021beyond}.  
In this work, we have designed an universal   TWA for dissipative quantum systems. Its
  simplicity, combined with its accuracy in solving key many-body quantum optics models, and low numerical cost, cf. Appendix~\ref{app:complexity}, strongly suggests its potential as an essential tool in every AMO physicist’s theoretical toolbox. Given the straightforward transition from the Lindblad master equation to the semi-classical equations of motion, developing a dedicated numerical framework to further streamline its use, similar to QuTiP for exact dynamics~\cite{Qutip} and QuantumCumulants.jl for CE~\cite{plankensteiner2022quantumcumulants}, is completely within reach. Such a framework could, in principle, leverage parallel computing techniques to efficiently explore problems in the actual many-body limit, encompassing thousands of degrees of freedom.

 This work can be extended by incorporating higher-order quantum corrections into the approximation. In this regard, our field-theoretic derivation provides a foundation for systematically advancing the semi-classical expansion. A similar approach has been previously explored for TWA in isolated systems~\cite{polkovnikov2010phase}, where next-order quantum corrections were introduced via quantum jumps, distinct from Lindblad jump operators, which are stochastically applied to trajectories. Including higher order corrections in our case follows the same logic. The connection between the truncated Wigner approximation (TWA) and the semi-classical limit of the Langevin equation lies at the heart of the transparency and flexibility of the approach presented here. 

Another possibility is to revisit the fundamental origins of dissipation. Specifically, when the environment is structured or non-Markovian, it modifies both the dissipative and stochastic components of the spin dynamics in~\eqref{eq:HSfield} (cf. Refs.~\cite{gardiner1985handbook,PhysRevA.37.4419,Breuer_nonMarkov2016}). This effect can be systematically derived using the same Keldysh path integral formalism employed in this work. In this framework, the damping term of dissipation is replaced by a convolution of dynamical variables with a memory function, $\Gamma_{ij}(t-t')$, while white noise is replaced by colored noise. Consequently, spin dynamics is governed by a set of stochastic integro-differential equations with temporally non-local deterministic terms, which are still amenable to efficient numerical techniques. This approach would provide a powerful tool for studying spin relaxation in structured environments, such as nitrogen-vacancy (NV) centers in diamond \cite{PhysRevB.106.L081122,Tserkovnyak,PhysRevLett.121.187204,PhysRevB.105.174412} or superconducting qubits~\cite{vonLupke_SCQubit2020}.

\begin{acknowledgments}
	The authors thank M. Stefanini and A. Gambassi for fruitful discussions. JM thanks M. Fleischhauer for insightful comments on a previous version of the manuscript. 
    This project has been supported by the Deutsche Forschungsgemeinschaft (DFG, German Research Foundation): through Project-ID 429529648, TRR 306 QuCoLiMa (``Quantum Cooperativity of Light and Matter'')  and
	by the QuantERA II Programme that has received funding from the European Union’s Horizon 2020 research and innovation programme under Grant Agreement No 101017733 ``QuSiED''); and by the Dynamics and Topology Center funded by the State of Rhineland Palatinate.  
	The authors gratefully acknowledge the computing time granted on the supercomputer MOGON 2 at Johannes Gutenberg-University Mainz (hpc.uni-mainz.de).  This research was supported in part by grant NSF PHY-2309135 to the Kavli Institute for Theoretical Physics (KITP).

\end{acknowledgments}

\appendix

\section{Derivation of semi-classical equations of motion}\label{app:derivation}

In this Appendix, we systematically obtain  the semi-classical equations of motion for dissipative spins, which were presented in the main text. Our approach, based on the Keldysh formalism of quantum field theory, is a natural extension of a similar derivation for unitary spin dynamics~\cite{Polkovnikov_2003,polkovnikov2010phase}.

Keldysh field theory is an alternative path-integral representation of quantum mechanics, and is particularly advantageous for non-equilibrium systems.~\cite{Kamenev_2023}. The conventional Feynman path-integral, whose lower and upper limits  respectively correspond to the initial ($t_i$) and final ($t_f$) times of the evolution, is given by
\begin{equation}
	S_\mathrm{Feynman} = \int_{t_i}^{t_f} L[\psi(t)]\,dt,
\end{equation}
where $\psi$ represents the degrees of freedom in the system. Instead, the Keldysh path-integral is defined along a closed time-contour $\mathcal{C}$:
\begin{equation}
	S_\mathrm{Keldysh} = \int_\mathcal{C} L[\psi(t_c)]\,dt_c, 
\end{equation}
where $\mathcal{C}$ is specified by the path $t_i \to t_f \to t_i$. Therefore, we can write
\begin{equation}
	S_\mathrm{Keldysh} =\int_{t_i}^{t_f} L[\psi^+(t)]\,dt - \int_{t_f}^{t_i} L[\psi^-(t)]\,dt,
\end{equation}
which means that one can work with a normal temporal integration at the expense of doubling the number of fields, corresponding to integration along the forward ($+$) and backward ($-$) branches of $\mathcal{C}$. The expectation values of operators can be obtained from
\begin{equation}\label{eq:expval_O}
	\expval{O(t)} = \int \mathbf{D}[\psi] \, P_0[\psi_0] \, O(t) \, e^{iS_\mathrm{Keldysh}},
\end{equation}
where $P_0[\psi]$ is the initial distribution of the fields. The advantage of working with a doubled number of fields is the ability to address non-equilibrium phenomena, in a wide range of problems from condensed matter physics to high energy physics and cosmology~\cite{Kamenev_2023}. Recently, it has been extended to open quantum systems with Lindblad dynamics~\cite{sieberer2016keldysh}, allowing the study of, for instance, the dissipative Dicke model~\cite{dalla2013keldysh,PhysRevA.87.023831} as well as the discovery of non-equilibrium universality classes~\cite{sieberer2023universality,Jamir_universality}.

Below, we present a step-by-step explanation of our derivation. Our discussion remains general, without specifying a particular dissipation channel. Moreover, we maintain an exact treatment throughout the derivation, applying the semi-classical approximation only at the final stage.

\subsection{Exact path-integral formulation of the problem}

Below, we apply a series of transformations to the Keldysh action of the original model, mapping the problem onto a field theory with explicit noise terms. In the next section, we will use this new formulation to derive the rules of dissipative TWA.
~\\

\emph{Constructing the Keldysh action--} The Keldysh action consists of a coherent part, corresponding to the Hamiltonian dynamics, and a dissipative part, corresponding to the Lindblad dynamics:
\begin{equation}
	S = S_c + S_d.
\end{equation}
The coherent part of the action can be written as
\begin{equation}\label{eq:S_c}
	S_c = S_0 + S_H,
\end{equation}
where $S_0$ is determined by the algebra of the degrees of freedom in the system and contains time-derivatives~\cite{Altland_Simons_2010}. $S_H$ is explicitly given by the Hamiltonian as
\begin{equation}\label{eq:S_H}
	S_H = -\int dt \,(H^+ - H^-),
\end{equation}
where $H^{\pm}$ is the Hamiltonian evaluated on the forward and backward branches of the Keldysh contour. The dissipative part of the action, corresponding to the Lindbladian in Eq.~\eqref{eq:LindbladMaster}, is given by (see Ref.~\cite{sieberer2016keldysh})
\begin{equation}\label{eq:S_d}
	S_d=-i \sum_i \frac{\gamma_i}{2}\int \mathrm{d} t\left(2 L_i^{+} \bar{L}_i^{-}-L_i^{-} \bar{L}_i^{-}-L_i^{+} \bar{L}_i^{+}\right)
\end{equation}
where $L_i$ are the fields corresponding to the jump operators. Without loss of generality, we have considered a diagonal matrix $\Gamma_{ij}=\delta_{ij}\gamma_i$, since the dissipation matrix can always be written in diagonal form after a linear transformation of jump operators.

~\\

\emph{`Recovering' the bath--} For technical reasons that will become evident later, we perform a procedure akin to recovering the bath degrees of freedom—in essence, the inverse of integrating out the bath. However, no additional information about the bath is needed beyond what is provided in Eq.~\eqref{eq:S_d}. Mathematically, this is accomplished through a Hubbard-Stratonovich transformation which reverses the process of evaluating a Gaussian integral~\cite{Altland_Simons_2010,Kamenev_2023}:
\begin{equation}\label{eq:HS_formula}
	e^{-\boldsymbol{u}^\dagger\cdot \hat{A} \cdot \boldsymbol{v}} \propto \int \mathbf{D}[\boldsymbol{x}^\dagger,\boldsymbol{x}] \,e^{-\boldsymbol{x}^\dagger \cdot \hat{A}^{-1}\cdot \boldsymbol{x}+i\boldsymbol{x}^\dagger\cdot \boldsymbol{v} + i \boldsymbol{u}^\dagger \cdot \boldsymbol{x}}.
\end{equation}
 We write the dissipative action in Eq.~\eqref{eq:S_d} in the matrix form
\begin{equation}
	S_d = \sum_i \int dt \, \begin{pmatrix}
		\bar{L}^+_i & -\bar{L}^-_i
	\end{pmatrix} \begin{pmatrix}
		i\gamma_i/2 & 0 \\ i\gamma_i & i \gamma_i/2
	\end{pmatrix}\begin{pmatrix}
		L^+_i \\ -L^-_i
	\end{pmatrix}.
\end{equation}
We then use Eq.~\eqref{eq:HS_formula} to get $e^{iS_d} = \int \boldsymbol{D}[\bar{\Phi}^\pm,\Phi^\pm]\,  e^{iS_B+iS_{sB}}$, where
\begin{equation}\label{eq:S_bath}
	S_B = \sum_i \int dt \,\begin{pmatrix}
		\bar{\Phi}^+_i & \bar{\Phi}^-_i
	\end{pmatrix} \begin{pmatrix}
		2i/\gamma_i & 0 \\ -4i/\gamma_i & 2i/\gamma_i
	\end{pmatrix}\begin{pmatrix}
		\Phi^+_i \\ \Phi^-_i
	\end{pmatrix},
\end{equation}
is the action of Markovian bath and 
\begin{equation}\label{eq:S_sbath}
	S_{sB} = \sum_i \int dt\, \big(\bar{\Phi}^+_i L^+_i + \bar{L}^+_i \Phi^+_i - \bar{\Phi}^-_i L^-_i - \bar{L}^-_i \Phi^-_i \big),
\end{equation}
describes the system-bath coupling. Together, Eqs.~\eqref{eq:S_c},~\eqref{eq:S_bath} and \eqref{eq:S_sbath} govern the dynamics of a system of spins coupled to the `fictitious' bath degrees of freedom, which are expressed by the fields $\Phi_i$. 
~\\

\emph{Separating quantum fluctuations--} The next step involves separating the fluctuating part of the fields due to quantum effects. This is done by defining the forward and backward fields in terms of classical and quantum fields, using the following linear transformation~\cite{Kamenev_2023}
\begin{align}\label{eq:Keldysh_pm_cq}
	\Phi^\pm_i &= \Phi^c_i \pm \Phi^q_i,\\
	L^\pm_i &= L^c_i \pm L^q_i.
\end{align} 
The action of the bath in the new basis is given by
\begin{equation}\label{eq:S_bath_qc}
	S_B = \sum_i \int \, \begin{pmatrix}
	    \bar{\Phi}^c_i & \bar{\Phi}^q_i
	\end{pmatrix}\begin{pmatrix}
	    0 & -4i/\gamma_i \\ 4i\gamma_i & 8i/\gamma_i
	\end{pmatrix}\begin{pmatrix}
	    \Phi^c_i \\ \Phi^q_i
	\end{pmatrix}\,dt,
\end{equation}
and the system-bath coupling reads as
\begin{equation}
	S_{sB} =  \sum_i \int dt\, \big(2\bar{\Phi}^c_i L^q_i + 2\bar{\Phi}^q_i L^c_i + 2\bar{L}^c_i \Phi^q_i + 2\bar{L}^q_i \Phi^c_i \big).
\end{equation}
We note that, the change of basis to the classical and quantum components  naturally leads to the Wigner transformation of the initial distribution function in Eq.~\eqref{eq:expval_O}~\cite{Polkovnikov_2003,polkovnikov2010phase}.

It is worth remarking that Eq.~\eqref{eq:S_bath_qc} can be directly extended to the non-Markovian regime, which is beyond the reach of Lindblad formalism but still amenable to our semiclassical approach.  To do so, one replaces the local kernel with the following generalized form
\begin{equation}
    S_B = \sum_i \iint \, \begin{pmatrix}
	    \bar{\Phi}^c_i(t) & \bar{\Phi}^q_i(t)
	\end{pmatrix}\cdot\hat{\bm{D}}^{-1}_i(t-t')\cdot\begin{pmatrix}
	    \Phi^c_i(t') \\ \Phi^q_i(t')
	\end{pmatrix}\,dt,
\end{equation}
where
\begin{equation}
   \hat{\bm{D}}_i(t-t') \equiv \begin{pmatrix}
	    0 & D^R_i(t-t') \\ D^A_i(t-t') & D_i^K(t-t')
	\end{pmatrix}
\end{equation}
is the Green's function of the environment in terms retarded and symmetric two-point correlation functions~\cite{Kamenev_2023}
\begin{align}
    D^R_i(t-t')&\equiv -\frac{i}{2}\Theta(t-t')\expval{[\hat{\Phi}_i(t),\hat{\Phi}_i(t')]},\\
    D^K_i(t-t')&\equiv -\frac{i}{2} \expval{\{\hat{\Phi}_i(t),\hat{\Phi}_i(t')\}},
\end{align}
and $ D^A_i(t-t')=\big(D^R_i)^*(t'-t)$. The Markovian regime can then be regarded as the limit where the correlation time of $\hat{\bm{D}}_i(t)$ becomes much smaller than the timescales of the system. This connection makes it clear that {the resulting stochastic equations (Eq.~\eqref{eq:C_Langevin}) have to be solved using the Stratonovich regularization of the noise.}
~\\

\emph{Identifying the noise--} In the next step, we identify and isolate the noisy contribution to dissipative dynamics. This is step similar to the derivation of the Langevin's equation~\cite{Kamenev_2023}, and is achieved by noting that the $\bar{\Phi}^q \Phi^q$ term in Eq.~\eqref{eq:S_bath_qc} can be written as
\begin{equation}\label{eq:qq_to_noise}
	e^{-\sum_i \frac{8}{\gamma_i} \int |\Phi^q_i|^2 \,dt} \propto \overline{e^{\sum_i \frac{2i}{\gamma_i} \int dt\, (\bar{\Phi}^q_i\xi_i + \bar{\xi}_i\Phi^q_i)\,dt}},
\end{equation}
where the average is taken with respect to the complex Gaussian noise $\xi(t)$ given by the following distribution
\begin{equation}\label{eq:noise_dist}
	\mathcal{P}[\bar{\xi}_i,\xi_i] \propto \exp(-\sum_i \frac{1}{2\gamma_i} \int |\xi_i(t)|^2\,dt).
\end{equation}
After substituting Eq.~\eqref{eq:qq_to_noise} in Eq.~\eqref{eq:S_bath_qc}, we get
\begin{multline}\label{eq:SB_plus_SsB}
	S_B + S_{sB} = \sum_i \int dt\, \Big[ \bar{\Phi}^q_i \Big( \frac{4i}{\gamma_i} \Phi^c_i+\frac{2}{\gamma_i}\xi_i  + 2{L}^c_i \Big) + \mathrm{c.c.} \Big] \\ + \sum_i \int dt \, \Big( 2\bar{\Phi}^c_i L^q_i + 2 \bar{L}^q_i\Phi^c_i \Big).
\end{multline}
Note that $\Phi^q$ only appears linearly in the first line. As the result, the functional integral over $(\bar{\Phi}^q,\Phi^q)$ can be evaluated, yielding a Dirac delta:
\begin{multline}
	\int \mathbf{D}[\bar{\Phi}^q_i,\Phi^q_i]e^{i\sum_i \int dt\, \Big[ \bar{\Phi}^q_i \big( \frac{4i}{\gamma_i} \Phi^c_i+\frac{2}{\gamma_i}\xi_i  + 2{L}^c_i \big) + \mathrm{c.c.} \Big]}\\ \propto \delta\Big(\frac{4i}{\gamma_i} \Phi^c_i+\frac{2}{\gamma_i}\xi_i  + 2\bar{L}^c_i \Big).
\end{multline}
Since $\Phi^q$ has been integrated out and we are left only with the classical field $\Phi^c$, for brevity, we ignore its classical index and define
\begin{equation}
	\Phi_i \equiv -i \Phi^c_i.
\end{equation}
The delta function determines the value of $\Phi_i$ as
\begin{equation}\label{eq:Phi_def}
	\Phi_i = \frac{\gamma_i}{2}L^c_i + \frac12 \xi_i.
\end{equation}
We rewrite the jump fields in the second line of Eq.~\eqref{eq:SB_plus_SsB} in the contour basis by using $2L^q_i = L^+_i - L^-_i$. Then, these terms can be absorbed into the Hamiltonian contribution in Eq.~\eqref{eq:S_H} to get
\begin{equation}
	\tilde{S}_H = -\int dt \, (\tilde{H}^+ - \tilde{H}^-),
\end{equation}
where $\tilde{H}$ is the complex effective Hamiltonian, which is $H$ shifted by a coupling to the jump fields:
\begin{equation}\label{eq:H_tilde}
	\tilde{H}^\pm \equiv H^\pm - i \sum_i \big(\bar{L}^\pm_i \Phi_i - \bar{\Phi}_i L^\pm_i \big).
\end{equation}
One might be tempted to elevate Eq.~\eqref{eq:H_tilde} to operator level. However, this is not possible as the field $\Phi$ depends on the jump fields (Eq.~\eqref{eq:Phi_def}) on both of the Keldysh contours, so $\tilde{H}^+$ ($\tilde{H}^-$) 
contains fields on both forward and backward contours.

So far, our treatment has been exact, and the full solution of the problem still requires evaluating the above noisy field theory. However, the new formulation of the problem is very convenient for semi-classical approximations, as we will show below.

\subsection{Semi-classical approximation}
To obtain the semi-classical approximation, we  look at the whole Keldysh path-integral that we obtained in the previous section:

\begin{multline}\label{eq:path_int_quant}
	\int \mathbf{D}[\bar{\Phi},\Phi]\,\int \mathbf{D}[\psi^\pm] \, P_0[\psi^\pm_0] \, \\ \times \overline{\delta\left(\Phi_i-\frac{\gamma_i}{2}L^c_i-\frac12 \xi_i\right)\,e^{iS_0 + i \tilde{S}_H}},
\end{multline}
where the long bar is the average with respect to noise distribution in Eq.~\eqref{eq:noise_dist}, and we have kept the delta function, which imposes the value of $\Phi$, explicit. In this formulation, the semi-classical approximation is straightforward: for each realization of the noise, we apply TWA to the path integral and then, take the average over different noise realizations. The former step yields the classical equations of motion averaged over different trajectories~\cite{polkovnikov2010phase}:
\begin{multline}\label{eq:quat_to_twa}
	\int\mathbf{D}[\psi^\pm]\, P_0[\psi^\pm_0]\,e^{iS_0+i\tilde{S}_H}\\ \approx \int \mathbf{D}[\psi^c]\, P_0^c[\psi_0^c]\,\delta\big(\dot{\psi}^c - \{\psi^c,\tilde{H}^c\}_p\big),
\end{multline}
where $P^c_0$ is the initial semi-classical (quasi-)distribution function, and $\{A,B\}_p$ is the Poisson's bracket. It is important to emphasize that we have not made specific assumptions about the nature of the system's degrees of freedom. These can correspond to bosonic modes, such as photons or phonons, or spin degrees of freedom, which may be represented using spin coherent states, Holstein-Primakoff bosons, or Schwinger bosons, independent of the spin size. Crucially, the resulting classical equations of motion, when expressed in terms of classical spin variables, are independent of the chosen spin representation. For brevity, we drop the classical index of all the fields as their quantum components no longer appear in Eq.~\eqref{eq:quat_to_twa}. Therefore, Eq.~\eqref{eq:path_int_quant} is approximated as
\begin{multline}
	\int \mathbf{D}[\bar{\Phi},\Phi] \, \mathbf{D}[\psi]\, P_0^c[\psi_0]\, \\ \times \overline{\delta\big(\dot{\psi}-\{\psi,\tilde{H}\}_p\big) \,\delta \big(\Phi_i-\frac{\gamma_i}{2}L^c_i-\frac12 \xi_i \big)}.
\end{multline}
The key point to remember is that, the first delta function on the RHS of the above equation yields the classical equations of motion for each trajectory while treating the field $\Phi$ in $\tilde{H}$ as a number, rather than   an active dynamical variable. Afterwards, the second delta function imposes the value of $\Phi$ in the equations of motion. In other words,$\Phi$ should not be treated as a dynamical variable, and \emph{its Poisson bracket with any other variable must be assumed to vanish}, when we derive the classical equations of motion.

We also note that, in the case of spins, we never had to explicitly use their path-integral representation. While one could, in principle, adopt a specific spin representation from the outset, such as spin-coherent states or Schwinger bosons, the final result always reduces to the form of Eq.~\eqref{eq:path_int_quant}, yielding nothing beyond the classical equations of motion~\cite{polkovnikov2010phase}.
~\\

Based on the discussion above, we arrive at the following rules for dissipative TWA:
~\\
\begin{enumerate}
	\item Find the equations of motion for the following classical complex Hamiltonian
	\begin{equation}
		\tilde{H} = H - i \sum_i (\bar{L}_i \Phi_i - \bar{\Phi}_i L_i),
	\end{equation}
	\item Substitute the self-consistent field $\Phi$ in the equations of motion according to
	\begin{equation}
		\Phi_i = \frac12 \gamma_i L_i + \frac12 \xi_i,
	\end{equation}
	where $\xi_i$ is a Gaussian noise defined by
	\begin{equation}
		\overline{\xi_i(t)}=0, \quad \overline{\xi_i(t) \bar{\xi}_i(t')} = 2\,\gamma_i \delta(t-t').
	\end{equation}
	\emph{Note that $\Phi_i$ has to be substituted only after obtaining the equations of motion.}
	\item Solve the equations of motion for different trajectories and different noise realizations until appropriate convergence is achieved.
\end{enumerate}
The extension to the general case of Eq.~\eqref{eq:LindbladMaster} with non-diagonal dissipation follows the same line of arguments, whose result was given in the text.

\section{Analytical estimate of quantum corrections}\label{app:error_estimate}

{TWA   fails to describe the correct steady state for the purely dissipative dynamics of sole spin loss ($\Omega=0$, Sec.~\ref{sec:driven_spin}). At a quantitative level, this can be appreciated by noting that while in the Heisenberg-Langevin approach the variance of the noise component for $\hat{\sigma}^z$ reads $\langle \hat{\xi}^z(t)\hat{\xi}^z(t')\rangle \propto\gamma_{\downarrow} (1+\hat{\sigma}^z)\delta(t-t')$, with $\hat{\xi}^z=-\xi_{\downarrow}^x \hat{\sigma}^x-\xi_{\downarrow}^y \hat{\sigma}^y$  (cf. Eqs~\eqref{eq:Lang}), in the  TWA  it reads instead:   $\langle 
\xi^z(t)\xi^z(t')\rangle\propto \gamma_{\downarrow}(s^2-(s^z)^2)\delta(t-t')$ (cf. Eqs.~\eqref{eq:EOM_single_z}).  The  steady state of   pure spin loss dynamics points to the south of the Bloch sphere, which is a `kernel' of the noise along the $z$ component, in   the Heisenberg-Langevin   dynamics. This does not happen   for the noise derived from the TWA. In fact, the TWA noise can never vanish for any physical spin configuration (recall that $s^2={3}$), implying that when there is no drive mitigating its effect, the TWA noise is always an over-estimate of the actual quantum noise. }
%

This observation calls for a quantitative estimate of  
the magnitude of quantum corrections to the dynamics of a driven spin, considered in Sec.~\ref{sec:driven_spin}. In order to do so, we consider a spin of size $S$ such that the norm of classical spins in each trajectory is given by $|\boldsymbol{s}|^2=4S^2$. The classical limit of this system is given by $S\to \infty$, while quantum fluctuations are strongest for $S=1/2$. Since TWA is a semi-classical approach, it naturally expands the dynamics around the classical limit. In order to obtain well-defined results in the classical limit, we rescale the decay rate according to $\gamma_\downarrow \to \gamma_\downarrow/2S$. Working with normalized spin variables defined as $\boldsymbol{\sigma} = \boldsymbol{s}/2S$, from Eqs.~\eqref{eq:EOM_single_x}-\eqref{eq:EOM_single_z} we obtain
\begin{align}
	\dv{}{t}\sigma^x &= \frac{\gamma_\downarrow}{2} \sigma^x \sigma^z + \xi^x \sigma^z,\label{eq:app_sigmax} \\
	\dv{}{t}\sigma^y &= -2\Omega \sigma^z + \frac{\gamma_\downarrow}{2} \sigma^y \sigma^z + \xi^y \sigma^z \\
	\dv{}{t}\sigma^z &= +2\Omega \sigma^y - \frac{\gamma_\downarrow}{2} \Big((\sigma^x)^2 + (\sigma^y)^2\Big)- \xi^x \sigma^x- \xi^y \sigma^y,
\end{align}
where the lack of explicit dependence on $S$ supports the rescaling of the decay with $S$. Modulo noise terms, these equations are identical to mean field equations for the decay of a collective spin, as used in the study of superradiant decay of atoms~\cite{Breuer_Petruccione}. The noise variances are given by
\begin{equation}
	\overline{\xi^\alpha(t) \xi^\beta(t')} = \frac{\gamma_\downarrow}{2S}\delta_{\alpha \beta}\delta(t-t').
\end{equation}
This already implies that \emph{the noise is suppressed for larger spins.} We also would like to understand how fluctuations are affected by the driving amplitude $\Omega$. In the main text, numerical data showed that the accuracy is higher for stronger drives, a fact which will analytically demonstrated below as well.


\begin{figure}[t!]
    \centering
    \includegraphics[width=0.8\linewidth]{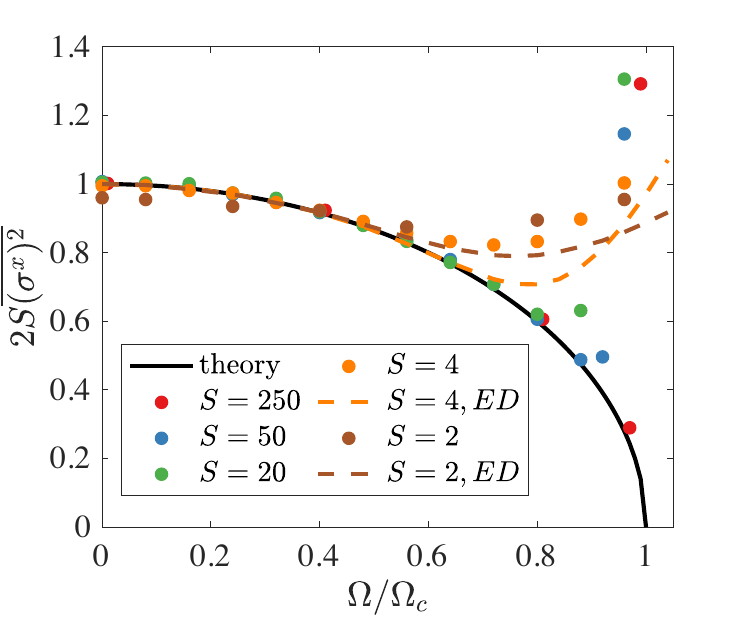}
    \caption{Noise correction to the variance of $\sigma^x$ for spins of size $S$    subject to coherent driving and incoherent decay. The numerical data (circles) agree with the analytical estimation (solid line) in Eq.~\eqref{eq:app_err_est} for small values of $\Omega/\Omega_c$ and large $S$. Dashed lines show exact solution for small system sizes. }
    \label{fig:app_err_est}
\end{figure}

Our strategy is to start from the fully classical limit, and incorporate the noise terms perturbatively. The classical system realizes a steady state given by $\sigma_0^x=0$ and
\begin{equation}
	\sigma^z_0 = - M \equiv \sqrt{1-\qty(\frac{\Omega}{\Omega_c})^2},
\end{equation}
where $\Omega_c$ is the critical Rabi frequency given by $\Omega_c=\gamma_\downarrow/4$. The above solution exists for $\Omega<\Omega_c$, otherwise the spin shows persisting oscillations. Assuming to be deep in the stationary limit, we approximately substitute the stationary value of $\sigma^z$ into Eq.~\eqref{eq:app_sigmax} and neglect its fluctuations. Solving the resulting equation, we find that
\begin{equation}
\lim_{t\to \infty} \sigma^x(t) = -Me^{-\gamma_\downarrow M t/2} \int_0^t e^{+\gamma_\downarrow M t'/2}\xi^x(t')\,dt'.
\end{equation}
Due to the noise having zero average, $\overline{\sigma^x(t)}$ vanishes and we have to calculate, at least, the second moment of $\sigma^x$ to resolve the contribution of the noise. We obtain
\begin{equation}\label{eq:app_err_est}
\overline{(\sigma^x)^2} = 0+ \underbrace{\frac{1}{2S}\sqrt{1-\qty(\frac{\Omega}{\Omega_c})^2}}_{\text{first order correction}} + \dots,
\end{equation}
where the zero term is the classical steady state value and dots represent higher order corrections. As shown in Fig.~\ref{fig:app_err_est}, Eq.~\eqref{eq:app_err_est} is consistent with   numerical results. At larger values of $\Omega$, we cannot substitute the steady-state value of $\sigma^z$ in the equation of motion for $\sigma^x$ as the fluctuations of the former become comparable to its expectation value. We remark that the deviation for $\Omega \to \Omega_c$ signals the failure of the analytical approximation used here, not TWA. Instead, the accuracy of TWA improves for larger values of $\Omega$, as was shown in the main text. Our result suggests that the expansion is in powers of $M/2S$, and higher order corrections become important as we reduce $\Omega$ below the critical value, or reduce the spin size. This is consistent with the numerical data given in the main text. We remark that for $S=1/2$  our result does not match the exact value $(\hat{\sigma}^x)^2=1$, as the expansion is made around the opposite extreme limit. Our analysis demonstrates the intricacy of estimating and characterizing quantum corrections in different systems, as can be seen from the non-trivial dependence of corrections on the physical parameters of the system, rather than a simple expansion in powers of $\hbar$ or $S$.

\section{Quantum corrections to multi-point correlation functions}\label{app:correlation}

As was mentioned in the text, the direct calculations of correlation functions using TWA only yields the fully symmetrized correlation functions. In other words, the result is blind to the non-trivial commutation relations between operators. Below, we demonstrate how to take these so-called quantum effects into account at the leading-order. The approach is identical to the one in Ref.~\cite{polkovnikov2010phase} for unitary dynamics, which is unaffected by the inclusion of Lindblad dissipation.

Before discussing more general cases, we consider the following correlation function
\begin{equation}\label{eq:def_C_AB}
    C_{AB}(t_A,t_B)= \langle\hat{A}(t_A)\hat{B}(t_B)\rangle,
\end{equation}
where $\hat{A}$ and $\hat{B}$ can be \emph{composite operators}, such that they are given by the product of the fundamental operators, i.e., those which directly appear in the equations of motion, according to
\begin{equation}
    \hat{A}=\hat{A}_{1}\hat{A}_{2}\dots \hat{A}_{n_A}
\end{equation}
\begin{equation}
    \hat{B}=\hat{B}_{1}\hat{B}_{2}\dots \hat{B}_{n_B}.
\end{equation}
For instance, $\hat{A}$ can be the photon number, such that
\begin{equation}
    \hat{A}_{1}=\hat{a}^\dagger, \quad \hat{A}_2 =\hat{a}.
\end{equation}
Taking $\tilde{A}$ and $\tilde{B}$ as the field-theoretic representation of these operators, we can write
\begin{equation}\label{eq:C_AB_fields}
    C_{AB}(t_A,t_B)=\langle \tilde{A}^-(t_A)\tilde{B}^{+}(t_B)\rangle,
\end{equation}
where putting $\tilde{A}$ ($\tilde{B}$) on the negative (positive) branch of the Keldysh contour ensures that it appears first (second) in the expectation value of operators. We note that the field-theoretic representations of operators do not necessarily match their operator forms. For example, if $\hat{A}=\hat{a}\hat{a}^\dagger$, then we need to normal-order it such that all of the annihilation operators appear on the right side of the operators~\cite{sieberer2016keldysh}. This can always be achieved by using commutations relations:
\begin{equation}
    \hat{a}\hat{a}^\dagger \xrightarrow{\textbf{normal \, order}}\hat{a}^\dagger\hat{a}+1.
\end{equation}
For spin operators, the best approach is to use commutation relations to reduce the power of spin operators to the smallest value possible, and then evaluate the field representation using spin coherent states. For example
\begin{equation}
    \bra{\boldsymbol{n}} (\hat{S}^\alpha)^2 \ket{\boldsymbol{n}}= \big(S^2-\frac{S}{2}\big)(\boldsymbol{n}\cdot \boldsymbol{e}_\alpha)^2 + \frac{S}{2}.
\end{equation}
For $S=1/2$, this simplifies to a number, in agreement with the algebra of Pauli matrices.

Having expressed the correlation function in terms of Keldysh fields, we proceed by decomposing them into their classical and quantum components (Eq.~\eqref{eq:Keldysh_pm_cq})
\begin{align}
    \tilde{A}^-(t_A)&=\tilde{A}^c(t_A)-\tilde{A}^q(t_A),\\
    \tilde{B}^+(t_B)&=\tilde{B}^c(t_B)+\tilde{B}^q(t_B).
\end{align}
Substitution in Eq.~\eqref{eq:C_AB_fields} gives
\begin{multline}
    C_{AB}(t_A,t_B)= \langle \tilde{A}^c(t_A)\tilde{B}^{c}(t_B)\rangle + \langle \tilde{A}^c(t_A)\tilde{B}^{q}(t_B)\rangle \\ - \langle \tilde{A}^q(t_A)\tilde{B}^{c}(t_B)\rangle - \langle \tilde{A}^q(t_A)\tilde{B}^{q}(t_B)\rangle.
\end{multline}
The first term is the symmetrized correlation function which can be directly obtained from TWA, and the last term vanishes as it only contains quantum fields~\cite{Kamenev_2023}. We need to evaluate the second and third terms. In the operator form, these correspond to
\begin{align}
    \langle \tilde{A}^c(t_A)\tilde{B}^{q}(t_B)\rangle &= \frac{i}{2}\chi_{AB}(t_A,t_B)\\
    \langle \tilde{A}^q(t_A)\tilde{B}^{c}(t_B)\rangle &= \frac{i}{2} \chi_{BA}(t_B,t_a),
\end{align}
where $\chi$ is the linear response function
\begin{equation}
    \chi_{AB}(t_A,t_B)=-i \Theta(t_A-t_B) \langle [ \hat{A}(t_A),\hat{B}(t_B) ]\rangle.
\end{equation}
According to the linear response theory~\cite{Altland_Simons_2010}, if we apply a small perturbation to the system which is coupled $\hat{B}$, as given by
\begin{equation}
    \delta H(t) = J(t) \hat{B},
\end{equation}
then, the change in $\langle \hat{A}\rangle$ for small values of $J$ to the leading-order is given by the Kubo formula~\cite{Altland_Simons_2010}
\begin{equation}
    \delta\langle \hat{A}(t_A)\rangle = \int_{-\infty}^{+\infty}\chi_{AB}(t_A,\tau) J(\tau)\,d\tau.
\end{equation}
Therefore, if $J(t)$ is a weak pulse centered at $\tau=t_B$
\begin{equation}
    J(\tau)=\delta(\tau-t_B),
\end{equation}
then
\begin{equation}
    \delta\langle \hat{A}(t_A)\rangle = - 2i\langle \tilde{A}^c(t_A)\tilde{B}^{q}(t_B)\rangle.
\end{equation}
In other words, by simulating the weak pulse in TWA and measuring the response, we can extract the leading order quantum corrections to correlation functions. This is also known as the method of quantum jumps~\cite{polkovnikov2010phase}, not to be mistaken with Lindblad jump operators.

For correlation functions of the form given by Eq.~\eqref{eq:def_C_AB}, it is always possible to assign fields to one of the two branches of the Keldysh contour and to carry out the semi-classical approximation as outlined above. If the number of temporal variables is more than two, then assigning fields to the different branches of the Keldysh contour is not possible in general, unless the operators appear in ``contour-ordered'' form in the correlation function. Notably, only correlation functions of this type appear naturally in the quantum theory of measurements~\cite{Nazarov_QuantumMeasurement2003,Clerk_QuantumMeasurement2010}. The treatment of correlation functions which cannot be cast in the contour-ordered format, such as the out-of-time-ordered correlators (OTOC), which are conceptually important but are not directly accessible in experiments, is beyond the scope of this work (cf. Ref.~\cite{Pappalardi_Echo} for a discussion of OTOCs using TWA).

\begin{table}[t]
\begin{tabular}{|l|l|l|}
\hline
Method                                                           & Number of equations                                                  & Accessible system size         \\ \hline
ED                                                               & $4^N$  ODE                                                           & $1-8$ spins                 \\ \hline
\begin{tabular}[c]{@{}l@{}}Stochastic \\ unraveling\end{tabular} & \begin{tabular}[c]{@{}l@{}}$2^N$ SDE \\ $N_{trj}$ times\end{tabular} & $1-12,\ldots, 16$ spins     \\ \hline
\begin{tabular}[c]{@{}l@{}} M'th order \\ cumulants \end{tabular}                                                       & $\propto N^M$ ODEs                                                    & $1-10^2$ spins               \\ \hline
TWA                                                              & \begin{tabular}[c]{@{}l@{}}$\propto N$ SDE \\ $N_{trj}$ times\end{tabular}  & $1-10^4,\ldots, 10^5$ spins \\ \hline
\end{tabular}
\caption{Comparison of numerical complexity of different methods for solving dissipative dynamics of   $N$ interacting spins-1/2.}\label{table:complexity}
\end{table}

\section{Numerical performance}\label{app:complexity}

In this Appendix, we briefly comment on the numerical complexity of different methods for solving dissipative dynamics of spin-1/2 systems (without relying on symmetries or permutational symmetry). A short overview of accessible system sizes to some of the commonly used methods is provided in Table~\ref{table:complexity}. Numerically exact methods are limited to small system sizes up to $16$ spins  (Table~\ref{table:complexity}). Semi-classical approaches, such as CE and TWA, can push the system size to larger values. Truncating CE at order $M$ requires solving a system of $\sim N^M$ differential equations, whose numerical costs grow quickly with the order of truncation and system size. However, the main challenge in using CE is the derivation of the equations of motion. This can become particularly cumbersome in the absence of symmetries which otherwise would force certain cumulants to vanish, simplifying the calculations. For instance, the symbolic evaluation of the equations of motion using the QuantumCumulants.jl package~\cite{plankensteiner2022quantumcumulants} takes more time than the numerical solution of the equations.

For TWA, one has to solve $3N$ stochastic differential equations of motion per each trajectory, which should be repeated $N_{trj}$ times, where the optimal value of $N_{trj} $ depends on the system size (for systems with periodic boundary conditions/permutation invariance one can also use an ensemble average to improve the precision of expectation values), the duration of time evolution (longer times require more trajectories in general), and even on the evaluated quantity. For instance, one point functions can converge for $10^3$ trajectories or less, while two-point functions may require $10^4-10^5$ trajectories. Nevertheless, the corresponding run-times are of the order of a few minutes on laptops, which, after taking into account the ease of implementation, makes TWA a viable tool for studying dissipative quantum systems. The run-time of TWA simulations scale linearly with the system size and therefore, there are no practical limitations for going to larger systems, since simulating multiple trajectories can be done using elementary parallel computation routines.


\bigskip 

\bibliography{SL}

\end{document}